\documentclass[aps,prb,showpacs,twocolumn,amsmath,amssymb,floatfix,superscriptaddress]{revtex4-1}
\usepackage{tabularx}
\usepackage{bm}
\usepackage{epsfig,subfigure}
\usepackage{multirow}
\usepackage{graphicx}
\usepackage{color}
\usepackage{amsfonts}
\usepackage{exscale}
\usepackage{amsbsy}
\usepackage[colorlinks,urlcolor=blue,linkcolor=blue,anchorcolor=blue,citecolor=blue,dvipdfm]{hyperref}

\begin{document}
\title{$\eta$-pairing in correlated fermion models with spin-orbit coupling}

\author{Kai Li}
\email[]{kaili@zzu.edu.cn}
\affiliation{School of Physics, Zhengzhou University, Zhengzhou 450001, China}
\affiliation{National Laboratory of Solid State Microstructures and School of Physics, Nanjing University, Nanjing 210093, China}

\begin{abstract}

We generalize the $\eta$-pairing theory in Hubbard models to the ones with spin-orbit coupling (SOC) and obtain the conditions under which the $\eta$-pairing operator is an eigenoperator of the Hamiltonian.
The $\eta$ pairing thus reveals an exact $SU(2)$ pseudospin symmetry in our spin-orbit coupled Hubbard model, even though the $SU(2)$ spin symmetry is explicitly broken by the SOC. In particular, these exact results can be applied to a variety of Hubbard models with SOC on either bipartite or non-bipartite lattices, whose noninteracting limit can be a Dirac semimetal, a Weyl semimetal, a nodal-line semimetal, and a Chern insulator. The $\eta$ pairing conditions also impose constraints on the band topology of these systems. We then construct and focus on an interacting Dirac-semimetal model, which exhibits an exact pseudospin symmetry with fine-tuned parameters. The stability regions for the \emph{exact} $\eta$-pairing ground states (with momentum $\bm{\pi}$ or $\bm{0}$) and the \emph{exact} charge-density-wave ground states are established. Between these distinct symmetry-breaking phases, there exists an exactly solvable multicritical line. In the end, we discuss possible experimental realizations of our results.

\end{abstract}
\date{\today}


\maketitle

\section{\label{sec:intro}introduction}

Exact results or solutions of interacting many-electron models beyond one spatial dimension are rare and highly appreciated. One important advance in studying the Hubbard model is the so-called $\eta$-pairing theory, which is an exact result due to Yang \cite{Yang1989}. The $\eta$-pairing operator allows for the construction of exact eigenstates possessing off-diagonal long-range order (ODLRO) and reveals an $SU(2)$ pseudospin symmetry \cite{Zhang1990} of the Hubbard model. More importantly, the $\eta$-pairing states can serve as exact superconducting ground states in a class of strongly correlated electronic models \cite{Essler1992,Essler1993,Shen1993,Boer1}, which is beyond the framework of BCS theory. Studies inspired by the $\eta$-pairing theory have since been pursued, including the unification of various symmetry-breaking orders (and related multicritical behavior study) \cite{Zhang1990,Zhang1997,Fradkin2015,Roy2018,Maciejko2020,Roy2020}, the construction of an $SU(2)$ gauge theory of the Hubbard model \cite{Kim2006,Kim2007,Hermele2007}, the calculation of entanglement entropy of $\eta$-pairing states \cite{Fan2005,Vafek2017}, and the study of quantum thermalization physics and quantum many-body scars in Hubbard-like models \cite{Nandkishore2015,Garrison2017,Veness2017,Clark2018,Bernevig2020,Motrunich2020}.

It is now recognized that spin-orbit coupling (SOC) can produce a wealth of topological phases and physical systems with SOC are currently under intense investigation. In particular, a number of theoretical understandings have been obtained for strongly correlated electronic systems with SOC \cite{PB2010,WCKB2014,Rau2016}. Nevertheless, most of these theoretical investigations used various approximate methods such as mean field theory and more rigorous and clearer physical understanding of these systems deserves further studies. Therefore, exact results or solutions for spin-orbit coupled interacting electronic systems would be valuable.

In this work, we first show that Yang's $\eta$-pairing theory can be generalized to a class of Hubbard models with SOC and derive the equations which make the $\eta$-pairing operator an eigenoperator of the Hamiltonian. The exact $SU(2)$ pseudospin symmetry can be then preserved in these systems, while the $SU(2)$ spin symmetry is explicitly broken by the SOC. We then exemplify these exact results with four concrete spin-orbit coupled systems, including a Dirac semimetal, a Weyl semimetal, a nodal-line semimetal, and a Chern insulator, in the noninteracting limit. Some consequences of the pseudospin symmetry and the relation between the present study and recent works are discussed. To convert the $\eta$-pairing eigenstates into ground states, additional interactions besides the Hubbard $U$ are needed. In particular, we construct and focus on an interacting Dirac-semimetal model on the square lattice.  By making the Hamiltonian frustration-free in some parameter regions, we establish the exact $\eta$-pairing ground states and the exact charge-density-wave (CDW) ground states, where the $\eta$-pairing ground states with momentum $\bm{\pi}$ and $\bm{0}$ are identified with a pair-density-wave (PDW) phase and a uniform superconductor, respectively. Moreover, there exists an exactly solvable multicritical line between the three symmetry-breaking phases. In the end, we discuss possible experimental realizations of our results in optical lattices.

This paper is organized as follows: In Sec.~\ref{sec:generalization} we introduce our model and generalize the $\eta$-pairing theory.
In Sec.~\ref{sec:application} we exemplify the generalized $\eta$-pairing theory with four concrete models.
An interacting Dirac-semimetal model with exact $\eta$-pairing ground states is presented in Sec.~\ref{sec:construction} and we identify an exactly solvable multicritical line in Sec.~\ref{sec:identification}.
We discuss possible experimental realizations in Sec.~\ref{sec:considerations} and present a summary and discussion in Sec.~\ref{sec:discussion}.
Certain details can be found in the appendices.

\section{Model and generalization of $\eta$ pairing\label{sec:generalization}}

The generic Hamiltonian for our spin-orbit coupled Hubbard models in arbitrary spatial dimensions is given by
\begin{eqnarray}
H&=&H_0+H_U
\nonumber\\
H_0&=&\sum_{\bm{k}}C_{\bm{k}}^\dag (\alpha_{\bm{k}}\sigma_x+\beta_{\bm{k}}\sigma_y+\gamma_{\bm{k}}\sigma_z+\varepsilon_{\bm{k}}\sigma_0)C_{\bm{k}}
\nonumber\\
H_U&=&U\sum_i n_{i\uparrow}n_{i\downarrow}.
\label{eq:SOU}
\end{eqnarray}
The noninteracting part $H_0$ is written in momentum space, where $C_{\bm{k}}^\dag=(c_{\bm{k}\uparrow}^\dag, c_{\bm{k}\downarrow}^\dag)$. In $H_0$, the three momentum-dependent real coefficients $\alpha_{\bm{k}}, \beta_{\bm{k}}, \gamma_{\bm{k}}$ with Pauli matrices $\sigma_{x,y,z}$ represent the SOC, and the term $\varepsilon_{\bm{k}}$ with the identity matrix $\sigma_0$ is the usual energy band without SOC. The interacting part $H_U$ written in real space is the on-site Hubbard interaction, where $n_{i\sigma}=c_{i\sigma}^\dagger c_{i\sigma}$.

In the absence of SOC ($\alpha_{\bm{k}}=\beta_{\bm{k}}=\gamma_{\bm{k}}=0$), Yang finds that the so-called $\eta$-pairing operator $\eta_{\bm{\pi}}^\dag=\sum_je^{i\bm{\pi}\cdot j}c_{j\uparrow}^\dag c_{j\downarrow}^\dag$ obeys the commutation relations $[H_0,\eta_{\bm{\pi}}^\dag]=-2\mu\eta_{\bm{\pi}}^\dag$ and $[H_U, \eta_{\bm{\pi}}^\dag]=U\eta_{\bm{\pi}}^\dag$, where $\mu$ is the chemical potential and $\bm{\pi}=(\pi,\pi,\pi)$ or $(\pi,\pi)$ for the cubic or square lattice. The total Hamiltonian and $\eta_{\bm{\pi}}^\dag$ thus obey $[H,\eta_{\bm{\pi}}^\dag]=(U-2\mu)\eta_{\bm{\pi}}^\dag$, and one can construct many exact eigenstates $H(\eta_{\bm{\pi}}^\dag)^m |0\rangle=m(U-2\mu)(\eta_{\bm{\pi}}^\dag)^m |0\rangle$ ($m=0,1,...$) possessing ODLRO.

In Yang's observation, the key point that makes the possible $\eta$ pairing is the fact that the energy band $\varepsilon_{\bm{k}}$ (for the cubic or square lattice without SOC) has the property $\varepsilon_{\bm{k}}+\varepsilon_{\bm{\pi}-\bm{k}}=-2\mu$ (\emph{independent of $\bm{k}$}), which leads to the commutation relation $[H_0,\eta_{\bm{\pi}}^\dag]=-2\mu\eta_{\bm{\pi}}^\dag$.

Therefore, to generalize the $\eta$-pairing theory to Hubbard models with SOC, e.g., Eq.\eqref{eq:SOU}, we need some new constraints on the coefficients $\alpha_{\bm{k}}, \beta_{\bm{k}}$, and $\gamma_{\bm{k}}$. To achieve these new constraints, we first define the $\eta$-pairing operator with a generic momentum $\bm{Q}$ as
\begin{equation}
\eta_{\bm{Q}}^\dag=\sum_{\bm{k}}c_{{\bm{k}}\uparrow}^\dag c_{\bm{Q}-{\bm{k}}\downarrow}^\dag=\sum_je^{i\bm{Q}\cdot j}c_{j\uparrow}^\dag c_{j\downarrow}^\dag.
\label{eq:etaQ}
\end{equation}
It can then be shown that (see Appendix \ref{eq4})
\begin{equation}
[H_0,\eta_{\bm{Q}}^\dag]=\lambda\eta_{\bm{Q}}^\dag,
\label{eq:H0eta}
\end{equation}
if $\alpha_{\bm{k}}, \beta_{\bm{k}}, \gamma_{\bm{k}}, \varepsilon_{\bm{k}}$, and $\bm{Q}$ satisfy the following equations
\begin{eqnarray}
\alpha_{\bm{k}}=\alpha_{\bm{Q}-\bm{k}} , \quad  \beta_{\bm{k}}=\beta_{\bm{Q}-\bm{k}},
\nonumber\\
\varepsilon_{\bm{k}}+\varepsilon_{\bm{Q}-\bm{k}}+\gamma_{\bm{k}}-\gamma_{\bm{Q}-\bm{k}}=\lambda ,
\label{eq:constraints}
\end{eqnarray}
where $\lambda$ is a constant \emph{independent of $\bm{k}$}. The commutator of the Hubbard interaction and the $\eta$-pairing operator reads
\begin{equation}
[H_U,\eta_{\bm{Q}}^\dag]=U\eta_{\bm{Q}}^\dag,
\label{eq:HUeta}
\end{equation}
which holds for arbitrary $\bm{Q}$. It follows from Eqs.\eqref{eq:H0eta} and \eqref{eq:HUeta} that the $\eta$-pairing operator is now an eigenoperator (note the different meaning of an eigenoperator introduced in the context of the rigged Hilbert space \cite{roberts1966a,roberts1966b}) of the total Hamiltonian, namely
\begin{equation}
[H,\eta_{\bm{Q}}^\dag]=(\lambda+U)\eta_{\bm{Q}}^\dag.
\label{eq:Heta}
\end{equation}
Therefore, one can construct many exact eigenstates $H(\eta_{\bm{Q}}^\dag)^m |0\rangle=m(\lambda+U)(\eta_{\bm{Q}}^\dag)^m |0\rangle$ possessing ODLRO.

The $SU(2)$ spin symmetry is explicitly broken by the SOC terms in Eq.\eqref{eq:SOU}. However, from Eq.\eqref{eq:Heta}, we see that our spin-orbit coupled Hubbard model possesses an exact $SU(2)$ \emph{pseudospin symmetry} when $\lambda+U=0$. The term pseudospin is attributed to the $SU(2)$ algebra formed by $\eta$-pairing operators, i.e., $[J_\alpha, J_\beta]=i\epsilon_{\alpha\beta\gamma}J_\gamma \ (\alpha,\beta, \gamma=x,y,z)$, where $J_x=\frac{1}{2}(\eta_{\bm{Q}}^\dag+\eta_{\bm{Q}}), J_y=\frac{1}{2i}(\eta_{\bm{Q}}^\dag-\eta_{\bm{Q}})$, $J_z=\frac{1}{2}(\hat{N}_e-N)$, $\hat{N}_e=\sum_{i\sigma}n_{i\sigma}$, and $N$ is the number of lattice sites.

Physically, the pseudospin symmetry can be understood via a particle-hole transformation for the spin-down electrons: $f_{j\uparrow}^\dag=c_{j\uparrow}^\dag, f_{j\downarrow}^\dag=e^{-i\bm{Q}\cdot j}c_{j\downarrow}$.
The pseudospin operators can now be rewritten in terms of the $f$ fermions as $J_\alpha=\frac{1}{2}\sum_jf_j^\dag\tau_\alpha f_j$, where $\tau_\alpha$ are the Pauli matrices and $f_j^\dag=(f_{j\uparrow}^\dag, f_{j\downarrow}^\dag)$. Thus, the pseudospin can be interpreted as the spin of $f$ fermions. This is reminiscent of the $SU(2)$ gauge structure in the fermionic-spinon representation of spin operators \cite{Affleck1988,Dagotto1988,Wen2002}. It is important to emphasize that the $\eta$-pairing states \emph{do not} have the pseudospin symmetry. In fact, they correspond to the ferromagnetic states of $f$ fermions with maximal total ''spin'' $N/2$, i.e., $\bm{J}^2(\eta_{\bm{Q}}^\dag)^m |0\rangle=\frac{N}{2}(\frac{N}{2}+1)(\eta_{\bm{Q}}^\dag)^m |0\rangle$ and $J_z(\eta_{\bm{Q}}^\dag)^m |0\rangle=(m-\frac{N}{2})(\eta_{\bm{Q}}^\dag)^m |0\rangle$ (where $\bm{J}^2=\sum_\alpha J_\alpha^2$), which break the $f$ fermion's spin symmetry and hence break the pseudospin symmetry.

\section{Application of the generalized $\eta$-pairing theory\label{sec:application}}

Based on the above exact results for model \eqref{eq:SOU}, we find that the the $\eta$ pairing can exist in a variety of spin-orbit coupled systems. In particular, we apply these exact results to four concrete examples, which include three gapless and one gapped noninteracting topological phases due to SOC.

The first example is a Dirac-semimetal Hubbard model on a square lattice. Its Hamiltonian is given by Eq.\eqref{eq:SOU} with $\alpha_{\bm{k}}=2t'\sin k_x, \beta_{\bm{k}}=2t'\sin k_y, \gamma_{\bm{k}}=0$, and $\varepsilon_{\bm{k}}=-2t(\cos k_x+\cos k_y)-\mu$. The lattice spacing has been set to unity throughout our paper. There are four Dirac points $\bm{K}=(n_1\pi,n_2\pi)$ in momentum space at $U=0$, where $n_{1,2}=0$ or $1$. Specifically, near the doubly degenerate point (e.g., $\bm{K}=(0,0)$) of the energy bands, the Bloch Hamiltonian $\mathcal{H}(\bm{k})=\alpha_{\bm{k}}\sigma_x+\beta_{\bm{k}}\sigma_y+\gamma_{\bm{k}}\sigma_z+\varepsilon_{\bm{k}}\sigma_0$ to the leading order reads $\mathcal{H}(\bm{k})\simeq2t'k_x\sigma_x+2t' k_y\sigma_y+(-4t-\mu)$, which is a 2-dimensional massless Dirac Hamiltonian \cite{Jackiw1984,Haldane1988,Melikyan2007}. Substituting the expressions of $\alpha_{\bm{k}}, \beta_{\bm{k}}, \gamma_{\bm{k}}$ and $\varepsilon_{\bm{k}}$ into Eq.\eqref{eq:constraints}, we find that the $\eta$-pairing momentum $\bm{Q}=\bm{\pi}$ and the constant $\lambda=-2\mu$.

Similarly, the next example we consider is a Weyl-semimetal Hubbard model on a cubic lattice \cite{William2014}. Its Hamiltonian is given by Eq.\eqref{eq:SOU} with $\alpha_{\bm{k}}=2t'\sin k_x, \beta_{\bm{k}}=2t'\sin k_y, \gamma_{\bm{k}}=2t'\sin k_z$, and $\varepsilon_{\bm{k}}=-2t(\cos k_x+\cos k_y+\cos k_z)-\mu$. There are totally eight Weyl points $\bm{K}=(n_1\pi,n_2\pi,n_3\pi)$ at $U=0$, where $n_{1,2,3}=0$ or $1$. As can be seen from Eq.\eqref{eq:constraints}, the $\eta$-pairing momentum $\bm{Q}=\bm{\pi}$ and the constant $\lambda=-2\mu$. Notice that for our Dirac- or Weyl-semimetal Hubbard model, the original $\eta$-pairing theory \cite{Yang1989} is recovered in the absence of SOC, e.g., $t'=0$.

Our third example is a nodal-line-semimetal Hubbard model on a cubic lattice \cite{Roy2017}. Its Hamiltonian is given by Eq.\eqref{eq:SOU} with $\alpha_{\bm{k}}=t(\cos k_x+\cos k_y-b), \beta_{\bm{k}}=t'\sin k_z, \gamma_{\bm{k}}=0$, and $\varepsilon_{\bm{k}}=-\mu$. At $U=0$, upon expanding $\alpha_{\bm{k}}$ and $\beta_{\bm{k}}$ near the $\bm{k}=\bm{0}$ point, we obtain a nodal ring with radius $\sqrt{2(2-b)}$ pinned at the $k_z=0$ plane. We notice that, from Eq.\eqref{eq:constraints}, the $\eta$-pairing momentum $\bm{Q}=(0,0,\pi)$ and the constant $\lambda=-2\mu$.

The last example we consider is a Chern-insulator Hubbard model on a triangular lattice \cite{Kai2016}. Its Hamiltonian is given by Eq.\eqref{eq:SOU} with $\alpha_{\bm{k}}=-2t\cos(\frac{k_x}{2}-\frac{\sqrt{3}k_y}{2}), \beta_{\bm{k}}=-2t\cos(\frac{k_x}{2}+\frac{\sqrt{3}k_y}{2}), \gamma_{\bm{k}}=-2t\cos k_x$, and $\varepsilon_{\bm{k}}=-\mu$. This model breaks time-reversal symmetry and $H_0$ has a nonzero band Chern number $\pm2$. In the large-$U$ limit, this system is described effectively by the triangular lattice Kitaev-Heisenberg spin model \cite{Kai2016,Kai2015}. We see that, from Eq.\eqref{eq:constraints}, the $\eta$-pairing momentum $\bm{Q}=\bm{0}$ and the constant $\lambda=-2\mu$. Note that the $\eta$-pairing operator with $\bm{Q}=\bm{0}$ is simply the $s$-wave Cooper-pairing operator in the usual BCS theory. In fact, the first generalization of $\eta$ pairing to a triangular lattice was obtained by adding a staggered $\pi/2$ flux through each triangle plaquette \cite{huizhai}, in the absence of SOC.

Now, a few remarks are in order. Firstly, as shown in Appendix \ref{topological}, the $\eta$ pairing conditions Eq.\eqref{eq:constraints} impose constraints on the topological properties of spin-orbit coupled systems. For example, the number of Dirac points in a 2-dimensional Dirac-semimetal must be a multiple of four, and the band Chern number of a Chern insulator must be an even integer. These have been embodied in the above concrete models.

Secondly, we notice that the $SU(2)$ pseudospin symmetry group contains a finite subgroup, namely the charge-conjugation symmetry  (see Appendix \ref{charge}), which leads to a vanishing $U(1)$ charge $\langle Q_i\rangle\equiv\langle n_i-1\rangle=0$ on each site, where $\langle A\rangle$ denotes the ensemble average of the operator $A$ at any temperature. For the above four examples of spin-orbit coupled Hubbard model \eqref{eq:SOU}, the exact $SU(2)$ pseudospin symmetry is respected when $\lambda+U=0$ or simply $\mu=U/2$, so the electron density is fixed at half filling (independent of the temperature) when $\mu=U/2$. Another consequence of the exact pseudospin symmetry is that the ODLRO and CDW long-range order must either be both absent or coexist \cite{Zhang1990,Shen1994}. In particular, the PDW phase (see below) possesses these two orders, whose parent Hamiltonian Eq.\eqref{eq:newH} exhibits an exact pseudospin symmetry (see Appendix \ref{eq7}).

Lastly, beyond these four concrete topological Hubbard models, the exact $SU(2)$ pseudospin symmetry has also been found in other Hubbard-like models, such as the extended Falicov-Kimball model \cite{Yunoki2019} and the Hubbard model with a Zeeman term \cite{Parameswaran2020}, which are in fact special examples of our general spin-orbit coupled Hubbard model \eqref{eq:SOU}. Therefore, the $\eta$ pairing structure and the associated $SU(2)$ pseudospin symmetry of the models in Refs.\cite{Yunoki2019,Parameswaran2020} can be readily obtained by using our generalized $\eta$-pairing theory for Hubbard models with SOC.

\section{Construction of the exact $\eta$-pairing ground states\label{sec:construction}}

In general, the exact $\eta$-pairing eigenstates of the Hubbard model \eqref{eq:SOU} are not ground states \cite{Yang1989}. In fact, beyond one spatial dimension, the Hubbard model has no exact solution to this day. However, some extended Hubbard models containing additional interactions are exactly solvable by constructing the ground state wave functions explicitly \cite{Essler1992,Essler1993,Boer1,Boer2}. Following the method used in Ref.\cite{Boer2}, we would further show that the $\eta$-pairing states can be constructed as the exact ground states of extended Hubbard models with SOC.

For the sake of concreteness, let us consider a square lattice Dirac-semimetal model similar to the one discussed above. Besides the on-site Hubbard $U$, we now introduce additional bond interactions, and the resulting new Hamiltonian can be expressed as
\begin{equation}
H'=\sum_{x\texttt{-bonds}}h_{ij}^x+\sum_{y\texttt{-bonds}}h_{ij}^y,
\label{eq:newH}
\end{equation}
where the summation $\sum_{x\texttt{-bonds}}$ ($\sum_{y\texttt{-bonds}}$) runs over all the bonds along the $x$ ($y$) direction of the square lattice. The bond Hamiltonians are given by
\begin{widetext}
\begin{eqnarray}
h_{ij}^x&=&t(c_i^\dag \sigma_x c_j+c_j^\dag \sigma_x c_i)-\frac{\mu}{4}(n_i+n_j)+\frac{U}{4}(n_{i\uparrow}n_{i\downarrow}+n_{j\uparrow}n_{j\downarrow})
\nonumber\\&&
+B\sum_\sigma(c_{i\sigma}^\dag c_{j\bar{\sigma}}+ c_{j\bar{\sigma}}^\dag c_{i\sigma})(n_{i\bar{\sigma}}+n_{j\sigma})+Vn_in_j+P(c_{i\uparrow}^\dag c_{i\downarrow}^\dag c_{j\downarrow}c_{j\uparrow}+c_{j\uparrow}^\dag c_{j\downarrow}^\dag c_{i\downarrow}c_{i\uparrow})
\nonumber\\
h_{ij}^y&=&t(c_i^\dag \sigma_y c_j+c_j^\dag \sigma_y c_i)-\frac{\mu}{4}(n_i+n_j)+\frac{U}{4}(n_{i\uparrow}n_{i\downarrow}+n_{j\uparrow}n_{j\downarrow})
\nonumber\\&&
+B\sum_\sigma(\bar{\sigma}ic_{i\sigma}^\dag c_{j\bar{\sigma}}+ \sigma ic_{j\bar{\sigma}}^\dag c_{i\sigma})(n_{i\bar{\sigma}}+n_{j\sigma})+Vn_in_j+P(c_{i\uparrow}^\dag c_{i\downarrow}^\dag c_{j\downarrow}c_{j\uparrow}+c_{j\uparrow}^\dag c_{j\downarrow}^\dag c_{i\downarrow}c_{i\uparrow}),
\label{eq:hij}
\end{eqnarray}
\end{widetext}
where $c_i^\dag=(c_{i\uparrow}^\dag, c_{i\downarrow}^\dag)$, $n_i=n_{i\uparrow}+n_{i\downarrow}$, and $\bar{\sigma}=-\sigma$. The noninteracting part $t$ represents SOC which is responsible for the Dirac points in momentum space. The $B$ term in Eq.\eqref{eq:hij} is known as the bond-charge interaction \cite{KSSH}, which completely breaks the $SU(2)$ spin-rotation symmetry and can be thought of as due to SOC \cite{Montorsi1991}. Note that it differs from the usual $SU(2)$ spin symmetric bond-charge interaction $\sum_\sigma(c_{i\sigma}^\dag c_{j\sigma}+ c_{j\sigma}^\dag c_{i\sigma})(n_{i\bar{\sigma}}+n_{j\bar{\sigma}})$ by spin flips on the bonds. In addition, the terms $V$ and $P$ denote the nearest-neighbor Coulomb interaction and the pair-hopping term, respectively.

We note that the $\eta$-pairing operator with momentum  $\bm{\pi}$ can be an eigenoperator of our extended Hubbard model with SOC, and hence Eq.\eqref{eq:newH} exhibits an exact pseudospin symmetry with fine-tuned parameters, as shown in Appendix \ref{eq7}. In the absence of the bond interactions ($B=V=P=0$), Eq.\eqref{eq:newH} reduces to the spin-orbit coupled Hubbard model \eqref{eq:SOU}, and one can show that, using the generalized $\eta$-pairing theory discussed above, the $\eta$-pairing operator with zero momentum is an eigenoperator.

To construct the exact ground states of Eq.\eqref{eq:newH}, the basic idea is to identify some parameter regions where the Hamiltonian \eqref{eq:newH} is \emph{frustration-free}: The ground states of Eq.\eqref{eq:newH} are simultaneous ground states of each and every local bond Hamiltonian $h_{ij}^{x, y}$ in Eq.\eqref{eq:hij}.

To find the parameter region (i.e., restrictions on $U$, $B$, $V$, and $P$ in the form of equalities and inequalities) where the $\eta$-pairing states are frustration-free ground states, let us diagonalize the local bond Hamiltonians \eqref{eq:hij}. Here, each bond Hamiltonian $h_{ij}^{x, y}$ has 16 local eigenstates and their respective energies, which are summarized in Table~\ref{tab:table}. Notice that we have set $B=-t$ in the calculation \cite{Essler1992,Essler1993,Boer1,Boer2}. For $B\neq-t$, the local eigenstates become complicated and it is difficult to see whether there exists a global frustration-free ground state \cite{Boer2}.

\begin{table}
\caption{\label{tab:table}This table summarizes the 16 local eigenstates and their respective eigenvalues of the bond Hamiltonians \eqref{eq:hij}, for $B=-t$. The 16 local bases are defined as follows: $|00\rangle$ denotes the vacuum, $|22\rangle=c_{i\uparrow}^\dag c_{i\downarrow}^\dag c_{j\uparrow}^\dag c_{j\downarrow}^\dag|00\rangle$, $|20\rangle=c_{i\uparrow}^\dag c_{i\downarrow}^\dag|00\rangle$, $|02\rangle=c_{j\uparrow}^\dag c_{j\downarrow}^\dag|00\rangle$, $|\sigma\sigma'\rangle=c_{i\sigma}^\dag c_{j\sigma'}^\dag|00\rangle$, $|\sigma0\rangle=c_{i\sigma}^\dag|00\rangle$, $|0\sigma\rangle= c_{j\sigma}^\dag|00\rangle$, $|\sigma2\rangle=c_{i\sigma}^\dag c_{j\uparrow}^\dag c_{j\downarrow}^\dag|00\rangle$, and $|2\sigma\rangle= c_{i\uparrow}^\dag c_{i\downarrow}^\dag c_{j\sigma}^\dag|00\rangle$.}
\begin{ruledtabular}
\begin{tabular}{ccc}
 &\multicolumn{1}{c}{$h_{ij}^x$}&\multicolumn{1}{c}{$h_{ij}^y$}\\
 Eigenvalue&Eigenstate&Eigenstate\\ \hline
 0&$|00\rangle$&$|00\rangle$ \\
 $\frac{U}{2}-\mu+4V$ & $|22\rangle$ & $|22\rangle$\\
 $\frac{U}{4}-\frac{\mu}{2}\pm P$ & $|20\rangle\pm|02\rangle$ & $|20\rangle\pm|02\rangle$\\
 $V-\frac{\mu}{2}$ & $|\sigma\sigma'\rangle$ & $|\sigma\sigma'\rangle$\\
 $\pm t-\frac{\mu}{4}$ & $|\sigma0\rangle\pm|0\bar{\sigma}\rangle$ & $|\sigma0\rangle\pm\sigma i|0\bar{\sigma}\rangle$\\
 $\pm t+\frac{U}{4}-\frac{3\mu}{4}+2V$ & $|\sigma2\rangle\mp|2\bar{\sigma}\rangle$ & $|\sigma2\rangle\mp\sigma i|2\bar{\sigma}\rangle$\\
\end{tabular}
\end{ruledtabular}
\end{table}

We note that the $\eta$-pairing states with momentum $\bm{\pi}=(\pi,\pi)$ , i.e., $(\eta_{\bm{\pi}}^\dag)^m |0\rangle$, can be built completely from the local states $|00\rangle$, $|22\rangle$, and $|20\rangle-|02\rangle$. Thus, $(\eta_{\bm{\pi}}^\dag)^m |0\rangle$ will be the common ground states of all the bond Hamiltonians $h_{ij}^x$ and $h_{ij}^y$, if $|00\rangle$, $|22\rangle$, and $|20\rangle-|02\rangle$ are local ground states. This requires that, from Table~\ref{tab:table}, $0=\frac{U}{2}-\mu+4V=\frac{U}{4}-\frac{\mu}{2}-P$ are the minimum eigenvalues, which yields the following constraints
\begin{eqnarray}
V&=&-\frac{P}{2}<0, \quad \mu =\frac{U}{2}+4V
\nonumber\\
U& <& \texttt{min}( -8|t|-8V,-4V).
\label{eq:constraints1}
\end{eqnarray}
Therefore, inside this parameter region together with $B=-t$, the exact ground states of our interacting Dirac-semimetal model \eqref{eq:newH} are $(\eta_{\bm{\pi}}^\dag)^m |0\rangle$ ($m=0,1,...,N$). The zero temperature ODLRO for this phase can be readily evaluated: $\langle c_{i\uparrow}^\dag c_{i\downarrow}^\dag c_{j\downarrow} c_{j\uparrow}\rangle_{T=0}=\sum_m\frac{\langle c_{i\uparrow}^\dag c_{i\downarrow}^\dag c_{j\downarrow} c_{j\uparrow}\rangle_m}{N+1}=\frac{1}{6}e^{i\bm{\pi}\cdot (j-i)}$, where $\langle c_{i\uparrow}^\dag c_{i\downarrow}^\dag c_{j\downarrow} c_{j\uparrow}\rangle_m=\frac{m(N-m)}{N(N-1)}e^{i\bm{\pi}\cdot (j-i)}$ is the ODLRO for each ground state $(\eta_{\bm{\pi}}^\dag)^m |0\rangle$ \cite{Yang1989}. It is this non-vanishing ODLRO $\langle c_{i\uparrow}^\dag c_{i\downarrow}^\dag c_{j\downarrow} c_{j\uparrow}\rangle_{T=0}$ at large distance $|i-j|\rightarrow\infty$ that characterizes the superconducting phase \cite{Yang1962} inside the region of Eq.\eqref{eq:constraints1}. This phase breaks the $U(1)$ particle-number symmetry, say, the number of electrons $2m$ in each ground state $(\eta_{\bm{\pi}}^\dag)^m |0\rangle$ is not fixed with changing $m$. In addition, the lattice translation symmetry is also broken due to the nonzero pairing momentum $\bm{\pi}$: Each ground state transforms as $(\eta_{\bm{\pi}}^\dag)^m |0\rangle\rightarrow(-1)^m(\eta_{\bm{\pi}}^\dag)^m |0\rangle$ under the translation by one lattice spacing along the $x$ or $y$ direction, but the factor $(-1)^m$ depends on the parity of $m$. We thus conclude that the above superconducting phase is in fact a PDW phase.


In the same way, one can establish the parameter region in which the $\eta$-pairing states with momentum $\bm{0}$ become the exact ground states. $(\eta_{\bm{0}}^\dag)^m |0\rangle$ is now built completely from the local states $|00\rangle$, $|22\rangle$, and $|20\rangle+|02\rangle$, which have to be made local ground states. Again from Table~\ref{tab:table}, we get the following constraints
\begin{eqnarray}
V&=&\frac{P}{2}<0, \quad \mu =\frac{U}{2}+4V
\nonumber\\
U& <& \texttt{min}( -8|t|-8V,-4V).
\label{eq:constraints2}
\end{eqnarray}
Similar to the $\eta$-pairing states with momentum $\bm{\pi}$, the phase inside this region also possesses a non-vanishing ODLRO $\langle c_{i\uparrow}^\dag c_{i\downarrow}^\dag c_{j\downarrow} c_{j\uparrow}\rangle_{T=0}=\frac{1}{6}$ and is thus a superconducting phase. The global $U(1)$ symmetry is spontaneously broken, due to the fluctuating electron numbers in the degenerate ground states $(\eta_{\bm{0}}^\dag)^m |0\rangle$. However, this phase preserves the lattice translation symmetry since the pairing momentum is $\bm{0}$. So it is a spatially uniform superconductor. In fact, as shown in Appendix \ref{swave}, it is closely related to the conventional $s$-wave superconductor.

It is interesting to see that the CDW state $|\texttt{CDW}\rangle=\prod_{i\in \mathcal{A}}c_{i\uparrow}^\dag c_{i\downarrow}^\dag|0\rangle$ can also be an exact ground state, where $\mathcal{A}$ denotes one sublattice of the square lattice. Notice that $|\texttt{CDW}\rangle$ can be constructed completely from the local states $|20\rangle$ and $|02\rangle$, we get the following constraints
\begin{eqnarray}
P&=&0
\nonumber\\
U& <& \texttt{min}( -4|t|+\mu,4V,2\mu)
\nonumber\\
V& >& \texttt{max}( \frac{|t|}{2}+\frac{\mu}{8},-\frac{U}{16}+\frac{\mu}{8}).
\label{eq:constraints3}
\end{eqnarray}
The CDW phase obviously breaks the lattice translation symmetry, accompanied by a charge density modulation with the same periodicity as that of the PDW phase. We notice that some important aspects of the physics of this simple CDW order in topological semimetals have been explored recently \cite{Burkov2020}.

Eqs.\eqref{eq:constraints1}-\eqref{eq:constraints3} are exact results that establish the stability regions of superconducting (i.e., the $\eta$-pairing states) and CDW ground states. We thus see that, for sufficiently small $U$, superconducting ground states are stabilized by attractive $V$ and finite $P$, whereas a repulsive $V$ (without $P$) favors a CDW ground state.

\section{Identification of an exact multicritical line\label{sec:identification}}

Given the three exactly solvable regions \eqref{eq:constraints1}-\eqref{eq:constraints3}, it is desirable to determine the phase transitions between these three symmetry-breaking phases. We note that the three regions \eqref{eq:constraints1}-\eqref{eq:constraints3} have \emph{no} overlaps with each other, however, overlaps exist in their boundaries. Specifically, as shown in Appendix \ref{eq12}, any two of the three boundaries of regions \eqref{eq:constraints1}-\eqref{eq:constraints3} share a common phase boundary, say
\begin{equation}
B=-t, V=P=0, U=2\mu\leq-8|t|.
\label{eq:multicritical}
\end{equation}
As a consequence, Eq.\eqref{eq:multicritical} is in fact an exact multicritical line, where the PDW, uniform superconductor, and CDW phases meet.

A complementary approach to confirm this is by checking the analyticity of ground-state energy near the line \eqref{eq:multicritical}. As can be seen from Table~\ref{tab:table}, the ground-state energy (per bond) vanishes continuously from each of the three phases toward the line \eqref{eq:multicritical}. However, its first derivatives (with respect to, e.g.,  $U$ and $\mu$) have a jump at line \eqref{eq:multicritical}, namely a kink, implying a quantum phase transition at this line.

More interestingly, the multicritical line \eqref{eq:multicritical} is also an exactly solvable line: Within this regime, we see that, from Table~\ref{tab:table}, the local ground states are $|00\rangle, |20\rangle, |02\rangle$, and $|22\rangle$. In this case, the ground-state space of Eq.\eqref{eq:newH} is spanned by the bases $\prod_i (c_{i\uparrow}^\dag c_{i\downarrow}^\dag)^{l_i}|0\rangle$ ($l_i=0,1$) and is highly degenerate, which contains both the $\eta$-pairing states (with arbitrary momentum) and the CDW states.


\section{Experimental considerations\label{sec:considerations}}

A few remarks are now in order concerning possible experimental realizations of our results in optical lattices. The essential ingredients of the present study are SOC and interaction. In general, the SOC can be implemented artificially in an optical lattice by introducing synthetic gauge fields \cite{Dalibard2011} and the interaction can be widely tuned through Feshbach resonances \cite{Chinrmp2010,Giorginirmp2008}. In particular, concrete schemes to realize spin-orbit coupled Hubbard models with controlled band structures and interactions have been proposed \cite{Duan2003,Zoller2004}. In fact, the Fermi-Hubbard model has been realized in an optical lattice \cite{Schneider2008,Jordens2008,Strohmaier2007}, where the kinetic energy, the interaction strength, and the chemical potential can be varied independently \cite{Hart2015,Mazurenko2017,Parsons2016,Boll2016,Cheuk2016,Brown2017}. Given these, it is plausible that the $SU(2)$ pseudospin symmetry can be realized via fine tuning of the model parameters, e.g., $\mu$ or $U$ so that $\mu=U/2$ as discussed previously, in cold-atom systems \cite{Bloch2008,Dalibard2011,Chinrmp2010,Giorginirmp2008}.

On the other hand, the exactly solvable interacting Dirac-semimetal model \eqref{eq:newH} contains off-site interactions, such as the $B$, $V$, and $P$ terms in Eq.\eqref{eq:hij}, which can arise from the matrix elements of the low-energy effective potential between fermionic atoms in an optical lattice \cite{Werner2005,Rosch2008}. Recent experiments in cold fermionic Rydberg atoms in an optical lattice have achieved a high degree of control over these interactions, such as adjusting the sign, strength, and range of the interactions \cite{Bloch2008,Sarma2015,Carroll2004,Viteau2011,Bloch2012,Labuhn2016,Chinrmp2010,Giorginirmp2008}. Therefore, the ability to control and tune interactions of extended Hubbard models in optical lattices has paved the way for the realization of Hamiltonian \eqref{eq:newH} over a wide range of parameters. Moreover, the $\eta$-pairing states with zero momentum are closely related to the $s$-wave superconductivity (see Appendix \ref{swave}), whose experimental evidence in an optical lattice has already been established and well studied \cite{Chin2006,Mitra2017}. We are therefore optimistic that the exact $\eta$-pairing ground states of Eq.\eqref{eq:newH} could be realized in optical lattices in the near future.

\section{Summary and discussion\label{sec:discussion}}

In summary, we have generalized Yang's $\eta$-pairing theory to a class of Hubbard models with SOC, where a hidden $SU(2)$ symmetry is uncovered. We apply these exact results to four concrete spin-orbit coupled Hubbard models, which include three gapless and one gapped topological phases in the noninteracting limit. We then construct and focus on an interacting Dirac-semimetal model and establish the stability regions for the \emph{exact} $\eta$-pairing ground states (with momentum $\bm{\pi}$ or $\bm{0}$) and the \emph{exact} CDW ground states. Between these distinct symmetry-breaking phases, there exists an exactly solvable multicritical line.

The present work may provide new guidance in exploring exotic quantum phases,  for instance a PDW phase here, in a wider range of physical systems, such as spin-orbit coupled systems with $SU(2)$ pseudospin symmetry. The exactly solvable interacting Dirac-semimetal model we constructed deserves further studies: This model hosts various \emph{exact} symmetry-breaking ground states and an exact multicritical line between them, which should serve as a clean platform to study intertwined orders \cite{Fradkin2015} and quantum phase transitions between the Dirac-semimetal and symmetry-breaking phases \cite{Classen2015,Roy2016,Assaad2017,Janssen2018,Maciejko2020,Torres2020}.

\begin{acknowledgments}
I am especially grateful to Jian-Xin Li and Zhenyu Zhang for helpful discussions and encouragement during the course of this work. I also thank Meng Cheng, Zhao-Long Gu, Zhao-Yang Dong, Feng Tang, Rui Wang, and Xiaoyu Zhu for enlightening conversations. This work was supported by the NSF of China under Grant No. 11704341 and the Open Research Program of the Laboratory of Solid State Microstructures (Nanjing University) under Grant No. M30020.
\end{acknowledgments}

\appendix

\section{Derivation of Eq.\eqref{eq:constraints} in the main text\label{eq4}}

Here, we show how to obtain Eq.\eqref{eq:constraints} in the main text. We first rewrite the noninteracting part $H_0$ in Eq.\eqref{eq:SOU} as
\begin{eqnarray}
H_0&=&\sum_{\bm{k}} [\alpha_{\bm{k}}(c_{\bm{k}\uparrow}^\dag c_{\bm{k}\downarrow}+c_{\bm{k}\downarrow}^\dag c_{\bm{k}\uparrow})+\beta_{\bm{k}}(-i c_{\bm{k}\uparrow}^\dag c_{\bm{k}\downarrow}+i c_{\bm{k}\downarrow}^\dag c_{\bm{k}\uparrow})
\nonumber\\&&
+(\varepsilon_{\bm{k}}+\gamma_{\bm{k}})c_{\bm{k}\uparrow}^\dag c_{\bm{k}\uparrow}+(\varepsilon_{\bm{k}}-\gamma_{\bm{k}})c_{\bm{k}\downarrow}^\dag c_{\bm{k}\downarrow}]
\label{eq:H0}.
\end{eqnarray}
It is now convenient to use the momentum space expression of $\eta$-pairing operator $\eta_{\bm{Q}}^\dag=\sum_{\bm{k}}c_{{\bm{k}}\uparrow}^\dag c_{\bm{Q}-{\bm{k}}\downarrow}^\dag$ to calculate its commutator with $H_0$.
After a direct and lengthy calculation, we get the following result
\begin{eqnarray}
[H_0,\eta_{\bm{Q}}^\dag]&=&\sum_{\bm{k}} [\alpha_{\bm{k}}c_{\bm{k}\downarrow}^\dag c_{\bm{Q}-{\bm{k}}\downarrow}^\dag-\alpha_{\bm{Q}-\bm{k}}c_{\bm{Q}-\bm{k}\uparrow}^\dag c_{\bm{k}\uparrow}^\dag
\nonumber\\&&
+i\beta_{\bm{k}}c_{\bm{k}\downarrow}^\dag c_{\bm{Q}-{\bm{k}}\downarrow}^\dag+i\beta_{\bm{Q}-\bm{k}}c_{\bm{Q}-\bm{k}\uparrow}^\dag c_{\bm{k}\uparrow}^\dag
\nonumber\\&&
+(\varepsilon_{\bm{k}}+\varepsilon_{\bm{Q}-\bm{k}}+\gamma_{\bm{k}}-\gamma_{\bm{Q}-\bm{k}})c_{{\bm{k}}\uparrow}^\dag c_{\bm{Q}-{\bm{k}}\downarrow}^\dag].
\label{eq:H0etaS}
\end{eqnarray}

Now, let us consider the first summation term $S=\sum_{\bm{k}} \alpha_{\bm{k}}c_{\bm{k}\downarrow}^\dag c_{\bm{Q}-{\bm{k}}\downarrow}^\dag$ in Eq.\eqref{eq:H0etaS}, and we have the following identities
\begin{eqnarray}
S&=&-\sum_{\bm{k}} \alpha_{\bm{k}} c_{\bm{Q}-{\bm{k}}\downarrow}^\dag c_{\bm{k}\downarrow}^\dag
\nonumber\\
S&=&\sum_{\bm{k'}} \alpha_{\bm{Q}-\bm{k'}}c_{\bm{Q}-\bm{k'}\downarrow}^\dag c_{{\bm{k'}}\downarrow}^\dag=\sum_{\bm{k}} \alpha_{\bm{Q}-\bm{k}}c_{\bm{Q}-\bm{k}\downarrow}^\dag c_{{\bm{k}}\downarrow}^\dag,
\label{eq:S}
\end{eqnarray}
where the first line follows from the anti-commutation relation of fermion operators, and the substitution $\bm{k}=\bm{Q}-\bm{k'}$ is used in the second line.  As can be seen from Eq.\eqref{eq:S}, if $\alpha_{\bm{k}}=\alpha_{\bm{Q}-\bm{k}}$, then we have $S=-S$ and hence $S=0$. The second summation term in Eq.\eqref{eq:H0etaS} also vanishes in the same way when $\alpha_{\bm{k}}=\alpha_{\bm{Q}-\bm{k}}$. Similarly, the third and the fourth summation terms in Eq.\eqref{eq:H0etaS} vanish separately when $\beta_{\bm{k}}=\beta_{\bm{Q}-\bm{k}}$. Finally, the last summation term in Eq.\eqref{eq:H0etaS} is equal to $\lambda\sum_{\bm{k}}c_{{\bm{k}}\uparrow}^\dag c_{\bm{Q}-{\bm{k}}\downarrow}^\dag=\lambda\eta_{\bm{Q}}^\dag$, if $\varepsilon_{\bm{k}}+\varepsilon_{\bm{Q}-\bm{k}}+\gamma_{\bm{k}}-\gamma_{\bm{Q}-\bm{k}}=\lambda$, where $\lambda$ is a constant \emph{independent of $\bm{k}$}.

In conclusion, when $\alpha_{\bm{k}}=\alpha_{\bm{Q}-\bm{k}}$, $\beta_{\bm{k}}=\beta_{\bm{Q}-\bm{k}}$, and $\varepsilon_{\bm{k}}+\varepsilon_{\bm{Q}-\bm{k}}+\gamma_{\bm{k}}-\gamma_{\bm{Q}-\bm{k}}=\lambda$, i.e., Eq.\eqref{eq:constraints}, we have $[H_0,\eta_{\bm{Q}}^\dag]=\lambda\eta_{\bm{Q}}^\dag$.

\section{Implications of $\eta$-pairing theory on the topological properties of spin-orbit coupled systems\label{topological}}

Based on our generalized $\eta$-pairing theory, in the following we will show that Eq.\eqref{eq:constraints} can have nontrivial restrictions on the properties of band topology of spin-orbit coupled systems.

Since we are interested in the topological properties, the usual energy dispersion term $\varepsilon_{\bm{k}}\sigma_0$ in $H_0$ can be ignored because it does not affect the band topology. Eq.\eqref{eq:constraints} is thus reduced to
\begin{equation}
\alpha_{\bm{k}}=\alpha_{\bm{Q}-\bm{k}}, \quad \beta_{\bm{k}}=\beta_{\bm{Q}-\bm{k}}, \quad \gamma_{\bm{k}}-\gamma_{\bm{Q}-\bm{k}}=\lambda=0,
\label{eq:constraint}
\end{equation}
where the last equation vanishes because $\lambda$ is independent of $\bm{k}$ and we can take $\bm{k}=\frac{\bm{Q}}{2}$ such that $\gamma_{\frac{\bm{Q}}{2}}-\gamma_{\bm{Q}-\frac{\bm{Q}}{2}}=\gamma_{\frac{\bm{Q}}{2}}-\gamma_{\frac{\bm{Q}}{2}}=0$.
Eq.\eqref{eq:constraint} implies that the corresponding Bloch Hamiltonian $\mathcal{H}(\bm{k})=\alpha_{\bm{k}}\sigma_x+\beta_{\bm{k}}\sigma_y+\gamma_{\bm{k}}\sigma_z$ has the property
\begin{equation}
\mathcal{H}(\bm{k})=\mathcal{H}(\bm{Q}-\bm{k}),
\label{eq:Hk}
\end{equation}
and the Bloch wave-functions also satisfy $\psi(\bm{k})=\psi(\bm{Q}-\bm{k})$ (up to an unphysical phase factor). From the point of view of topology, this implies that each point $\bm{k}$ is topologically identical to the point $\bm{Q}-\bm{k}$ in momentum space.

Let us now focus on 2-dimensional systems. If we translate the Brillouin zone (BZ) by vector $\frac{\bm{Q}}{2}$ and choose its center as the new origin of coordinate in momentum space, then the above arguments mean that each point $\bm{k}$ is topologically identical to the point $-\bm{k}$. We now bipartite the BZ into two halves, then each half BZ is topologically identical to a sphere which is a closed manifold (See Appendix A 3 in Ref.\cite{Kai2016} for the proof).

Therefore, we arrive at the following conclusions:

(1). \emph{If the spin-orbit coupled system is a 2-dimensional Dirac-semimetal, then the number of Dirac points must be a multiple of four}.

Proof: Firstly, the two half BZs host equal number of Dirac points due to the symmetry property of Bloch Hamiltonian, e.g., Eq.\eqref{eq:Hk}. Secondly, each half BZ is a closed manifold where the Dirac points must appear in pairs.

(2). \emph{If the system is a Chern insulator, then the band Chern number must be an even integer}. [See Appendix A 3 in Ref.\cite{Kai2016} for the proof.]

\section{The $\eta$-pairing states are topologically trivial\label{swave}}

In this section, we argue that the $\eta$-pairing states, e.g., $(\eta_{\bm{0}}^\dag)^m |0\rangle$ ($m=0,1,...,N$), are closely related to the conventional BCS superconductivity and hence they are topologically trivial. Specifically, let us consider the $s$-wave BCS mean-field Hamiltonian $H_{BCS}=\sum_{\bm{k}\sigma}\xi_{\bm{k}}c_{\bm{k}\sigma}^\dag c_{\bm{k}\sigma}+\sum_{\bm{k}}\Delta_{\bm{k}}(c_{{\bm{k}}\uparrow}^\dag c_{-{\bm{k}}\downarrow}^\dag+c_{{-\bm{k}}\downarrow} c_{{\bm{k}}\uparrow})$, where $\Delta_{\bm{k}}<0$. The intra-pair distance between the two electrons in the usual Cooper pair is finite, whereas it is zero (i.e., on-site singlet pairing) in the $\eta$-pairing states. Therefore, the ground state of $H_{BCS}$ is in general not a linear combination of $(\eta_{\bm{0}}^\dag)^m |0\rangle$.

However, $H_{BCS}$ is \emph{topologically} equivalent to a simpler Hamiltonian
\begin{equation}
H_{\eta}=\sum_{\bm{k}}\Delta_{\bm{k}}(c_{{\bm{k}}\uparrow}^\dag c_{-{\bm{k}}\downarrow}^\dag+c_{{-\bm{k}}\downarrow} c_{{\bm{k}}\uparrow}),
\end{equation}
since the energy spectrum remains fully gapped during the deformation $\xi_{\bm{k}}\rightarrow0$ and there should be no \emph{topological} phase transition. The ground state wave function (up to a normalization factor) of $H_{\eta}$ is now readily written as $|G\rangle=\prod_{\bm{k}}(1+c_{{\bm{k}}\uparrow}^\dag c_{-{\bm{k}}\downarrow}^\dag)|0\rangle=\sum_{m=0}^N\frac{1}{m!}(\eta_{\bm{0}}^\dag)^m |0\rangle$, which is a superposition of $\eta$-pairing states $(\eta_{\bm{0}}^\dag)^m |0\rangle$ \cite{Shirakawa2020}. Since $H_{BCS}$ is topologically trivial, $H_{\eta}$ and hence $(\eta_{\bm{0}}^\dag)^m |0\rangle$ are also topologically trivial.

An alternative way to see that the $\eta$-pairing states are topologically trivial may be as follows. As shown in the main text, the $\eta$-pairing states can be mapped to the ferromagnetic states through a \emph{local} particle-hole transformation. A local transformation should not change the global topological property of a state. Therefore, the $\eta$-pairing states are topologically trivial since the ferromagnetic states are topologically trivial.

\section{Charge-conjugation symmetry as a subgroup of the $SU(2)$ pseudospin symmetry\label{charge}}

As shown in the main text, the generators of the $SU(2)$ pseudospin symmetry group are the pseudospin operators $J_x,J_y$ and $J_z$. The global pseudospin-rotation by $\pi$ about the $J_y$ axis, say $\mathcal{C}=e^{i\pi J_y}$, flips the sign of $J_z$: $\mathcal{C}J_z\mathcal{C}^{-1}=-J_z$. We recall that the pseudospin operator $J_z$ is proportional to the $U(1)$ charge operator $\hat{Q}\equiv\hat{N}_e-N$, say $J_z=\frac{1}{2}\hat{Q}$. Therefore,  the pseudospin-rotation $\mathcal{C}=e^{i\pi J_y}$ is in fact a charge conjugation \cite{Hermele2007,You2017}, i.e.,
\begin{equation}
\mathcal{C}\hat{Q}\mathcal{C}^{-1}=-\hat{Q}.
\label{eq:charge}
\end{equation}
In fact, $\mathcal{C}=e^{i\pi J_y}$ flips the sign of the local operator $J_i^z\equiv\frac{1}{2}Q_i\equiv\frac{1}{2}(n_i-1)$ on each site. So, if the model Hamiltonian has the $SU(2)$ pseudospin symmetry (and hence the charge-conjugation symmetry), then $\langle Q_i\rangle=\langle J_i^z\rangle=0$ on each site, where $\langle A\rangle$ denotes the ensemble average of the operator $A$ at any temperature.

It is interesting to see that the charge-conjugation operator $\mathcal{C}$ squares to the charge parity \cite{You2017}
\begin{equation}
\mathcal{C}^2=(-1)^{\hat{Q}},
\label{eq:chargesquare1}
\end{equation}
since $\mathcal{C}^2=e^{i2\pi J_y}=e^{i2\pi J_z}=(-1)^{2J_z}$ and $J_z=\frac{1}{2}\hat{Q}$. An alternative way to understand Eq.\eqref{eq:chargesquare1} is to make an analogy between the charge-conjugation $\mathcal{C}$ and the ''time reversal'' $\mathcal{T}_f$. Specifically, let us focus on the $f$ fermions introduced by a particle-hole transformation for the spin-down electrons and we have $J_\alpha=\frac{1}{2}\sum_jf_j^\dag\tau_\alpha f_j$, as shown in the main text. The charge-conjugation $\mathcal{C}=e^{i\pi J_y}$ rotates the spin of $f$ fermions $\mathcal{C}f_{j\uparrow}\mathcal{C}^{-1}=-f_{j\downarrow}$, $\mathcal{C}f_{j\downarrow}\mathcal{C}^{-1}=f_{j\uparrow}$. The ''time reversal'' $\mathcal{T}_f$ acts on these $f$-fermion operators the same as $\mathcal{C}$, except that $\mathcal{T}_f$ is an antiunitary operator, e.g., $\mathcal{T}_fi\mathcal{T}_f^{-1}=-i$. So, we can define $\mathcal{T}_f$ to be
\begin{equation}
\mathcal{T}_f=\mathcal{C}\mathcal{K},
\label{eq:timecharge}
\end{equation}
where $\mathcal{K}$ denotes the complex conjugation operator. To be precise, consider the $f$-fermion occupation number basis $\prod_{j=1}^N (f_{j\uparrow}^\dag)^{l_{j\uparrow}} (f_{j\downarrow}^\dag)^{l_{j\downarrow}} |0\rangle$, where $|0\rangle$ is the vacuum state of $f$ fermions and $l_{j\sigma}\in\{0,1\}$. Now the operator $\mathcal{K}$ is defined in this basis, e.g., the base kets $\prod_{j=1}^N (f_{j\uparrow}^\dag)^{l_{j\uparrow}} (f_{j\downarrow}^\dag)^{l_{j\downarrow}} |0\rangle$ are invariant under the action of $\mathcal{K}$. In this representation, the charge-conjugation $\mathcal{C}=e^{i\pi J_y}$ acts as
\begin{equation}
\mathcal{C}=\left(\begin{array}{cc} i\tau_y & 0 \\ 0 & \mathbf{I} \\\end{array} \right)^{\otimes N},
\label{eq:chargematrix}
\end{equation}
where $\tau_y$ is the second Pauli matrix and $\mathbf{I}$ is the $2\times2$ identity matrix. The unitary operator $\mathcal{C}$ thus turns out to be a \emph{real} orthogonal matrix in the above representation, say $\mathcal{C}^{\ast}=\mathcal{C}$. Using Eq.\eqref{eq:timecharge}, we have $\mathcal{T}_f^{2}=\mathcal{C}\mathcal{C}^\ast=\mathcal{C}^2$. Now we recall that the ''time reversal'' operator $\mathcal{T}_f$ squares to the fermion number parity $\mathcal{T}_{f}^2=(-1)^{\hat{N}_f}$, with $\hat{N}_f=\sum_{j,\sigma}f_{j\sigma}^\dag f_{j\sigma}$. We finally conclude that
\begin{equation}
\mathcal{C}^2=(-1)^{\hat{N}_f}.
\label{eq:chargesquare2}
\end{equation}
Note that Eq.\eqref{eq:chargesquare2} is indeed equivalent to Eq.\eqref{eq:chargesquare1}, since $\hat{Q}$ and $\hat{N}_f$ have the same parity: $\hat{N}_f=2J_z+2\sum_j f_{j\downarrow}^\dag f_{j\downarrow}=\hat{Q}+2\sum_j f_{j\downarrow}^\dag f_{j\downarrow}$, where the term $2\sum_j f_{j\downarrow}^\dag f_{j\downarrow}$ always has an even parity.

For the sake of completeness, we outline the action of the charge-conjugation operator $\mathcal{C}$ on the original electron operators, say
\begin{equation}
\mathcal{C}c_{j\sigma}\mathcal{C}^{-1}=-\sigma e^{i\bm{Q}\cdot j}c_{j,-\sigma}^\dag,
\label{eq:chargeaction}
\end{equation}
for which reason $\mathcal{C}$ is also called particle-hole transformation.

\section{Implications of $\eta$-pairing theory on the superconducting properties in Hubbard models with SOC\label{superconducting}}

Following Zhang's idea in Ref.\cite{Zhang1990}, we will show that the $SU(2)$ pseudospin symmetry of our spin-orbit coupled Hubbard models would lead to a pair of new collective modes together with the usual Goldstone mode, if the system is in the superconducting phase.

The crucial point in Zhang's observation is the fact that the transverse components of the pseudospin operators look like a superconducting order, while the $z$ component looks like a CDW order, up to a local phase factor $e^{i\bm{Q}\cdot j}$. So the multiplet of the particle-particle and particle-hole operators should transform as a vector under the pseudospin rotation
\begin{equation}
[J_+,\Delta_z]=-2\Delta_+, \quad [J_-,\Delta_+]=-\Delta_z, \quad [J_z,\Delta_+]=\Delta_+,
\label{eq:SCCDW}
\end{equation}
meaning that the pseudospin operators can \emph{rotate} the superconducting order and the CDW order into each other, where the pseudospin operators $J_+=J_x+iJ_y=\eta_{\bm{Q}}^\dag$, $J_-=J_+^\dag$, $J_z=\frac{1}{2}(\hat{N}_e-N)$, the s-wave Cooper-pairing operator $\Delta_+=\sum_jc_{j\uparrow}^\dag c_{j\downarrow}^\dag$, $\Delta_-=\Delta_+^\dag$, and the CDW operator $\Delta_z=\sum_{j\sigma}e^{-i\bm{Q}\cdot j}n_{j\sigma}$ with the wave-vector $\bm{Q}$.

To study the collective modes, let us consider the following response function
\begin{equation}
D_+(t,t')=-\frac{i}{N}\Theta(t-t')\langle [J_+(t),\Delta_z(t')]\rangle.
\label{eq:D+t}
\end{equation}
Since the Hamiltonian is time-independent, we have $D_+(t,t')=D_+(t-t')=-\frac{i}{N}\Theta(t-t')\langle [J_+(t-t'),\Delta_z]\rangle$. From Eq.\eqref{eq:Heta} in the main text, the time dependence of pseudospin operators in the Heisenberg representation can be determined explicitly, in particular, $J_+(t-t')=e^{i(\lambda+U)(t-t')}J_+$. Using the commutator in Eq.\eqref{eq:SCCDW}, Eq.\eqref{eq:D+t} can be thus calculated exactly, i.e., $D_+(t-t')=2i\rho\Theta(t-t')e^{i(\lambda+U)(t-t')}$, where $\rho=\frac{\langle\Delta_+\rangle}{N}$ denotes the superconducting order parameter.

Information about collective modes are now embedded in the analytic properties of the Fourier transform of Eq.\eqref{eq:D+t}, say
\begin{equation}
D_+(\omega)=\frac{-2\rho}{\omega+(\lambda+U)+i\delta},
\label{eq:D+}
\end{equation}
where $\delta$ is a positive infinitesimal. Similarly, we have
\begin{eqnarray}
D_-(t,t')&=&-\frac{i}{N}\Theta(t-t')\langle [J_-(t),\Delta_z^\dag(t')]\rangle
\nonumber\\
D_-(\omega)&=&\frac{2\rho^*}{\omega-(\lambda+U)+i\delta},
\label{eq:D-}
\end{eqnarray}
and
\begin{eqnarray}
D_z(t,t')&=&-\frac{i}{N}\Theta(t-t')\langle [J_z(t),\Delta_+(t')]\rangle
\nonumber\\
D_z(\omega)&=&\frac{\rho}{\omega+i\delta}.
\label{eq:Dz}
\end{eqnarray}

Therefore, as can be seen from Eqs.\eqref{eq:D+}, \eqref{eq:D-}, and \eqref{eq:Dz}, if the ground state of our spin-orbit coupled Hubbard model is in the superconducting phase (i.e., $\rho\neq0$), then there must exist a triplet of collective modes with energies $-(\lambda+U)$, $\lambda+U$, and 0, respectively. Note that the collective mode with energy 0 in Eq.\eqref{eq:Dz} is simply the gapless Goldstone mode resulting from the spontaneous $U(1)$ symmetry breaking of superconductivity. In addition, we find that there are a pair of new collective modes with energies $\pm(\lambda+U)$, which is also a consequence of superconductivity.

\section{The $SU(2)$ pseudospin symmetry of Eq.\eqref{eq:newH} in the main text\label{eq7}}

As shown in the main text, the $\eta$-pairing operator can be an eigenoperator of a class of Hubbard models with SOC. In fact, the $\eta$-pairing operator can also be an eigenoperator of an extended Hubbard model with SOC, such as the interacting Dirac-semimetal model Eq.\eqref{eq:newH} we studied in the main text.

To see this, we calculate the commutator of the Hamiltonian \eqref{eq:newH} and the $\eta$-pairing operator $\eta_{\bm{\pi}}^\dag$, which reads
\begin{eqnarray}
[H',\eta_{\bm{\pi}}^\dag]&=&-2(t+B)\sum_{x\texttt{-bonds}}e^{i\bm{\pi}\cdot j}(c_{i\uparrow}^\dag c_{j\uparrow}^\dag-c_{i\downarrow}^\dag c_{j\downarrow}^\dag)
\nonumber\\&&
+2i(t+B)\sum_{y\texttt{-bonds}}e^{i\bm{\pi}\cdot j}(c_{i\uparrow}^\dag c_{j\uparrow}^\dag+c_{i\downarrow}^\dag c_{j\downarrow}^\dag)
\nonumber\\&&
+(2V+P)\sum_{\langle i,j\rangle}e^{i\bm{\pi}\cdot j}(c_{j\uparrow}^\dag c_{j\downarrow}^\dag n_i-c_{i\uparrow}^\dag c_{i\downarrow}^\dag n_j)
\nonumber\\&&
+(U-2\mu-4P)\eta_{\bm{\pi}}^\dag.
\label{eq:newHeta}
\end{eqnarray}
Therefore, the $\eta$-pairing operator $\eta_{\bm{\pi}}^\dag$ becomes an eigenoperator of $H'$ and $(\eta_{\bm{\pi}}^\dag)^m |0\rangle$ become exact eigenstates with eigenvalues $m(U-2\mu-4P)$, when $B=-t, V=-\frac{P}{2}$. Moreover, if $B=-t, V=-\frac{P}{2}$, and $\mu=\frac{U}{2}+4V$, $H'$ commutes with $\eta_{\bm{\pi}}^\dag$ and hence possesses the $SU(2)$ pseudospin symmetry. Accordingly, within the regime of the PDW phase where $(\eta_{\bm{\pi}}^\dag)^m |0\rangle$ are the exact ground states (see Eq.\eqref{eq:constraints1} in the main text), the corresponding Hamiltonian \eqref{eq:newH} exhibits an exact pseudospin symmetry which gets spontaneously broken.

\begin{widetext}

\section{Derivation of Eq.\eqref{eq:multicritical} in the main text\label{eq12}}

Eq.\eqref{eq:multicritical} describes the common phase boundary shared by the three exactly solvable regions \eqref{eq:constraints1}, \eqref{eq:constraints2}, and \eqref{eq:constraints3}, which respectively host the PDW, uniform superconductor, and CDW phases. To obtain Eq.\eqref{eq:multicritical}, we need first to work out the boundary of each region.

From Eq.\eqref{eq:constraints1}, we see that its boundary $\partial_\texttt{PDW}$ consists of two disjoint sets
\begin{equation}
\partial_\texttt{PDW}=\{V=P=0, U=2\mu\leq-8|t|\}\cup\{V=-\frac{P}{2}<0, \mu =\frac{U}{2}+4V, U=\texttt{min}( -8|t|-8V,-4V)\}.
\label{eq:boundary9}
\end{equation}

Similarly, the boundary of Eq.\eqref{eq:constraints2} reads
\begin{equation}
\partial_\texttt{SC}=\{V=P=0, U=2\mu\leq-8|t|\}\cup\{V=\frac{P}{2}<0, \mu =\frac{U}{2}+4V, U=\texttt{min}( -8|t|-8V,-4V)\}.
\label{eq:boundary10}
\end{equation}

Lastly, the boundary of Eq.\eqref{eq:constraints3} reads
\begin{eqnarray}
\partial_\texttt{CDW}&=&\{P=0, U\leq\texttt{min}( -4|t|+\mu,4V,2\mu), V=\texttt{max}( \frac{|t|}{2}+\frac{\mu}{8},-\frac{U}{16}+\frac{\mu}{8})\}
\nonumber\\&&
\cup\{P=0, U=\texttt{min}( -4|t|+\mu,4V,2\mu), V>\texttt{max}( \frac{|t|}{2}+\frac{\mu}{8},-\frac{U}{16}+\frac{\mu}{8})\}.
\label{eq:boundary11}
\end{eqnarray}
Note that we have omitted $B=-t$ in all of the above equations.

From these equations, we find that any two of the boundaries $\partial_\texttt{PDW}$, $\partial_\texttt{SC}$, and $\partial_\texttt{CDW}$ have equal intersections
\begin{equation}
\partial_\texttt{PDW}\cap\partial_\texttt{SC}=\partial_\texttt{SC}\cap\partial_\texttt{CDW}=\partial_\texttt{CDW}\cap\partial_\texttt{PDW}=\{B=-t, V=P=0, U=2\mu\leq-8|t|\}.
\label{eq:commonboundary}
\end{equation}
As a consequence, Eqs.\eqref{eq:constraints1},\eqref{eq:constraints2},\eqref{eq:constraints3} in the main text share a common phase boundary
\begin{equation}
\partial_\texttt{PDW}\cap\partial_\texttt{SC}\cap\partial_\texttt{CDW}=\{B=-t, V=P=0, U=2\mu\leq-8|t|\},
\label{eq:commonboundary1}
\end{equation}
which is a multicritical line between these distinct symmetry-breaking phases.
\end{widetext}

\bibliography{ref}

\begin{thebibliography}{77}%
\makeatletter
\providecommand \@ifxundefined [1]{%
 \@ifx{#1\undefined}
}%
\providecommand \@ifnum [1]{%
 \ifnum #1\expandafter \@firstoftwo
 \else \expandafter \@secondoftwo
 \fi
}%
\providecommand \@ifx [1]{%
 \ifx #1\expandafter \@firstoftwo
 \else \expandafter \@secondoftwo
 \fi
}%
\providecommand \natexlab [1]{#1}%
\providecommand \enquote  [1]{``#1''}%
\providecommand \bibnamefont  [1]{#1}%
\providecommand \bibfnamefont [1]{#1}%
\providecommand \citenamefont [1]{#1}%
\providecommand \href@noop [0]{\@secondoftwo}%
\providecommand \href [0]{\begingroup \@sanitize@url \@href}%
\providecommand \@href[1]{\@@startlink{#1}\@@href}%
\providecommand \@@href[1]{\endgroup#1\@@endlink}%
\providecommand \@sanitize@url [0]{\catcode `\\12\catcode `\$12\catcode
  `\&12\catcode `\#12\catcode `\^12\catcode `\_12\catcode `\%12\relax}%
\providecommand \@@startlink[1]{}%
\providecommand \@@endlink[0]{}%
\providecommand \url  [0]{\begingroup\@sanitize@url \@url }%
\providecommand \@url [1]{\endgroup\@href {#1}{\urlprefix }}%
\providecommand \urlprefix  [0]{URL }%
\providecommand \Eprint [0]{\href }%
\providecommand \doibase [0]{http://dx.doi.org/}%
\providecommand \selectlanguage [0]{\@gobble}%
\providecommand \bibinfo  [0]{\@secondoftwo}%
\providecommand \bibfield  [0]{\@secondoftwo}%
\providecommand \translation [1]{[#1]}%
\providecommand \BibitemOpen [0]{}%
\providecommand \bibitemStop [0]{}%
\providecommand \bibitemNoStop [0]{.\EOS\space}%
\providecommand \EOS [0]{\spacefactor3000\relax}%
\providecommand \BibitemShut  [1]{\csname bibitem#1\endcsname}%
\let\auto@bib@innerbib\@empty
\bibitem [{\citenamefont {Yang}(1989)}]{Yang1989}%
  \BibitemOpen
  \bibfield  {author} {\bibinfo {author} {\bibfnamefont {C.~N.}\ \bibnamefont
  {Yang}},\ }\href {\doibase 10.1103/PhysRevLett.63.2144} {\bibfield  {journal}
  {\bibinfo  {journal} {Phys. Rev. Lett.}\ }\textbf {\bibinfo {volume} {63}},\
  \bibinfo {pages} {2144} (\bibinfo {year} {1989})}\BibitemShut {NoStop}%
\bibitem [{\citenamefont {Zhang}(1990)}]{Zhang1990}%
  \BibitemOpen
  \bibfield  {author} {\bibinfo {author} {\bibfnamefont {S.}~\bibnamefont
  {Zhang}},\ }\href {\doibase 10.1103/PhysRevLett.65.120} {\bibfield  {journal}
  {\bibinfo  {journal} {Phys. Rev. Lett.}\ }\textbf {\bibinfo {volume} {65}},\
  \bibinfo {pages} {120} (\bibinfo {year} {1990})}\BibitemShut {NoStop}%
\bibitem [{\citenamefont {Essler}\ \emph {et~al.}(1992)\citenamefont {Essler},
  \citenamefont {Korepin},\ and\ \citenamefont {Schoutens}}]{Essler1992}%
  \BibitemOpen
  \bibfield  {author} {\bibinfo {author} {\bibfnamefont {F.~H.~L.}\
  \bibnamefont {Essler}}, \bibinfo {author} {\bibfnamefont {V.~E.}\
  \bibnamefont {Korepin}}, \ and\ \bibinfo {author} {\bibfnamefont
  {K.}~\bibnamefont {Schoutens}},\ }\href {\doibase
  10.1103/PhysRevLett.68.2960} {\bibfield  {journal} {\bibinfo  {journal}
  {Phys. Rev. Lett.}\ }\textbf {\bibinfo {volume} {68}},\ \bibinfo {pages}
  {2960} (\bibinfo {year} {1992})}\BibitemShut {NoStop}%
\bibitem [{\citenamefont {Essler}\ \emph {et~al.}(1993)\citenamefont {Essler},
  \citenamefont {Korepin},\ and\ \citenamefont {Schoutens}}]{Essler1993}%
  \BibitemOpen
  \bibfield  {author} {\bibinfo {author} {\bibfnamefont {F.~H.~L.}\
  \bibnamefont {Essler}}, \bibinfo {author} {\bibfnamefont {V.~E.}\
  \bibnamefont {Korepin}}, \ and\ \bibinfo {author} {\bibfnamefont
  {K.}~\bibnamefont {Schoutens}},\ }\href {\doibase 10.1103/PhysRevLett.70.73}
  {\bibfield  {journal} {\bibinfo  {journal} {Phys. Rev. Lett.}\ }\textbf
  {\bibinfo {volume} {70}},\ \bibinfo {pages} {73} (\bibinfo {year}
  {1993})}\BibitemShut {NoStop}%
\bibitem [{\citenamefont {Shen}\ and\ \citenamefont {Qiu}(1993)}]{Shen1993}%
  \BibitemOpen
  \bibfield  {author} {\bibinfo {author} {\bibfnamefont {S.-Q.}\ \bibnamefont
  {Shen}}\ and\ \bibinfo {author} {\bibfnamefont {Z.-M.}\ \bibnamefont {Qiu}},\
  }\href {\doibase 10.1103/PhysRevLett.71.4238} {\bibfield  {journal} {\bibinfo
   {journal} {Phys. Rev. Lett.}\ }\textbf {\bibinfo {volume} {71}},\ \bibinfo
  {pages} {4238} (\bibinfo {year} {1993})}\BibitemShut {NoStop}%
\bibitem [{\citenamefont {de~Boer}\ \emph {et~al.}(1995)\citenamefont
  {de~Boer}, \citenamefont {Korepin},\ and\ \citenamefont
  {Schadschneider}}]{Boer1}%
  \BibitemOpen
  \bibfield  {author} {\bibinfo {author} {\bibfnamefont {J.}~\bibnamefont
  {de~Boer}}, \bibinfo {author} {\bibfnamefont {V.~E.}\ \bibnamefont
  {Korepin}}, \ and\ \bibinfo {author} {\bibfnamefont {A.}~\bibnamefont
  {Schadschneider}},\ }\href {\doibase 10.1103/PhysRevLett.74.789} {\bibfield
  {journal} {\bibinfo  {journal} {Phys. Rev. Lett.}\ }\textbf {\bibinfo
  {volume} {74}},\ \bibinfo {pages} {789} (\bibinfo {year} {1995})}\BibitemShut
  {NoStop}%
\bibitem [{\citenamefont {Zhang}(1997)}]{Zhang1997}%
  \BibitemOpen
  \bibfield  {author} {\bibinfo {author} {\bibfnamefont {S.-C.}\ \bibnamefont
  {Zhang}},\ }\href {\doibase 10.1126/science.275.5303.1089} {\bibfield
  {journal} {\bibinfo  {journal} {Science}\ }\textbf {\bibinfo {volume}
  {275}},\ \bibinfo {pages} {1089} (\bibinfo {year} {1997})}\BibitemShut
  {NoStop}%
\bibitem [{\citenamefont {Fradkin}\ \emph {et~al.}(2015)\citenamefont
  {Fradkin}, \citenamefont {Kivelson},\ and\ \citenamefont
  {Tranquada}}]{Fradkin2015}%
  \BibitemOpen
  \bibfield  {author} {\bibinfo {author} {\bibfnamefont {E.}~\bibnamefont
  {Fradkin}}, \bibinfo {author} {\bibfnamefont {S.~A.}\ \bibnamefont
  {Kivelson}}, \ and\ \bibinfo {author} {\bibfnamefont {J.~M.}\ \bibnamefont
  {Tranquada}},\ }\href {\doibase 10.1103/RevModPhys.87.457} {\bibfield
  {journal} {\bibinfo  {journal} {Rev. Mod. Phys.}\ }\textbf {\bibinfo {volume}
  {87}},\ \bibinfo {pages} {457} (\bibinfo {year} {2015})}\BibitemShut
  {NoStop}%
\bibitem [{\citenamefont {Roy}\ and\ \citenamefont {Foster}(2018)}]{Roy2018}%
  \BibitemOpen
  \bibfield  {author} {\bibinfo {author} {\bibfnamefont {B.}~\bibnamefont
  {Roy}}\ and\ \bibinfo {author} {\bibfnamefont {M.~S.}\ \bibnamefont
  {Foster}},\ }\href {\doibase 10.1103/PhysRevX.8.011049} {\bibfield  {journal}
  {\bibinfo  {journal} {Phys. Rev. X}\ }\textbf {\bibinfo {volume} {8}},\
  \bibinfo {pages} {011049} (\bibinfo {year} {2018})}\BibitemShut {NoStop}%
\bibitem [{\citenamefont {Boyack}\ \emph {et~al.}(2020)\citenamefont {Boyack},
  \citenamefont {Yerzhakov},\ and\ \citenamefont {Maciejko}}]{Maciejko2020}%
  \BibitemOpen
  \bibfield  {author} {\bibinfo {author} {\bibfnamefont {R.}~\bibnamefont
  {Boyack}}, \bibinfo {author} {\bibfnamefont {H.}~\bibnamefont {Yerzhakov}}, \
  and\ \bibinfo {author} {\bibfnamefont {J.}~\bibnamefont {Maciejko}},\ }\href
  {https://arxiv.org/abs/2004.09414} {\bibfield  {journal} {\bibinfo  {journal}
  {arXiv:2004.09414}\ } (\bibinfo {year} {2020})}\BibitemShut {NoStop}%
\bibitem [{\citenamefont {Szabo}\ and\ \citenamefont {Roy}(2020)}]{Roy2020}%
  \BibitemOpen
  \bibfield  {author} {\bibinfo {author} {\bibfnamefont {A.}~\bibnamefont
  {Szabo}}\ and\ \bibinfo {author} {\bibfnamefont {B.}~\bibnamefont {Roy}},\
  }\href {https://arxiv.org/abs/2009.05055} {\bibfield  {journal} {\bibinfo
  {journal} {arXiv:2009.05055}\ } (\bibinfo {year} {2020})}\BibitemShut
  {NoStop}%
\bibitem [{\citenamefont {Kim}(2006)}]{Kim2006}%
  \BibitemOpen
  \bibfield  {author} {\bibinfo {author} {\bibfnamefont {K.-S.}\ \bibnamefont
  {Kim}},\ }\href {\doibase 10.1103/PhysRevLett.97.136402} {\bibfield
  {journal} {\bibinfo  {journal} {Phys. Rev. Lett.}\ }\textbf {\bibinfo
  {volume} {97}},\ \bibinfo {pages} {136402} (\bibinfo {year}
  {2006})}\BibitemShut {NoStop}%
\bibitem [{\citenamefont {Kim}(2007)}]{Kim2007}%
  \BibitemOpen
  \bibfield  {author} {\bibinfo {author} {\bibfnamefont {K.-S.}\ \bibnamefont
  {Kim}},\ }\href {\doibase 10.1103/PhysRevB.75.245105} {\bibfield  {journal}
  {\bibinfo  {journal} {Phys. Rev. B}\ }\textbf {\bibinfo {volume} {75}},\
  \bibinfo {pages} {245105} (\bibinfo {year} {2007})}\BibitemShut {NoStop}%
\bibitem [{\citenamefont {Hermele}(2007)}]{Hermele2007}%
  \BibitemOpen
  \bibfield  {author} {\bibinfo {author} {\bibfnamefont {M.}~\bibnamefont
  {Hermele}},\ }\href {\doibase 10.1103/PhysRevB.76.035125} {\bibfield
  {journal} {\bibinfo  {journal} {Phys. Rev. B}\ }\textbf {\bibinfo {volume}
  {76}},\ \bibinfo {pages} {035125} (\bibinfo {year} {2007})}\BibitemShut
  {NoStop}%
\bibitem [{\citenamefont {Fan}\ and\ \citenamefont {Lloyd}(2005)}]{Fan2005}%
  \BibitemOpen
  \bibfield  {author} {\bibinfo {author} {\bibfnamefont {H.}~\bibnamefont
  {Fan}}\ and\ \bibinfo {author} {\bibfnamefont {S.}~\bibnamefont {Lloyd}},\
  }\href {http://stacks.iop.org/0305-4470/38/i=23/a=014} {\bibfield  {journal}
  {\bibinfo  {journal} {Journal of Physics A: Mathematical and General}\
  }\textbf {\bibinfo {volume} {38}},\ \bibinfo {pages} {5285} (\bibinfo {year}
  {2005})}\BibitemShut {NoStop}%
\bibitem [{\citenamefont {Vafek}\ \emph {et~al.}(2017)\citenamefont {Vafek},
  \citenamefont {Regnault},\ and\ \citenamefont {Bernevig}}]{Vafek2017}%
  \BibitemOpen
  \bibfield  {author} {\bibinfo {author} {\bibfnamefont {O.}~\bibnamefont
  {Vafek}}, \bibinfo {author} {\bibfnamefont {N.}~\bibnamefont {Regnault}}, \
  and\ \bibinfo {author} {\bibfnamefont {B.~A.}\ \bibnamefont {Bernevig}},\
  }\href {\doibase 10.21468/SciPostPhys.3.6.043} {\bibfield  {journal}
  {\bibinfo  {journal} {SciPost Phys.}\ }\textbf {\bibinfo {volume} {3}},\
  \bibinfo {pages} {043} (\bibinfo {year} {2017})}\BibitemShut {NoStop}%
\bibitem [{\citenamefont {Nandkishore}\ and\ \citenamefont
  {Huse}(2015)}]{Nandkishore2015}%
  \BibitemOpen
  \bibfield  {author} {\bibinfo {author} {\bibfnamefont {R.}~\bibnamefont
  {Nandkishore}}\ and\ \bibinfo {author} {\bibfnamefont {D.~A.}\ \bibnamefont
  {Huse}},\ }\href {\doibase 10.1146/annurev-conmatphys-031214-014726}
  {\bibfield  {journal} {\bibinfo  {journal} {Annual Review of Condensed Matter
  Physics}\ }\textbf {\bibinfo {volume} {6}},\ \bibinfo {pages} {15} (\bibinfo
  {year} {2015})}\BibitemShut {NoStop}%
\bibitem [{\citenamefont {Garrison}\ \emph {et~al.}(2017)\citenamefont
  {Garrison}, \citenamefont {Mishmash},\ and\ \citenamefont
  {Fisher}}]{Garrison2017}%
  \BibitemOpen
  \bibfield  {author} {\bibinfo {author} {\bibfnamefont {J.~R.}\ \bibnamefont
  {Garrison}}, \bibinfo {author} {\bibfnamefont {R.~V.}\ \bibnamefont
  {Mishmash}}, \ and\ \bibinfo {author} {\bibfnamefont {M.~P.~A.}\ \bibnamefont
  {Fisher}},\ }\href {\doibase 10.1103/PhysRevB.95.054204} {\bibfield
  {journal} {\bibinfo  {journal} {Phys. Rev. B}\ }\textbf {\bibinfo {volume}
  {95}},\ \bibinfo {pages} {054204} (\bibinfo {year} {2017})}\BibitemShut
  {NoStop}%
\bibitem [{\citenamefont {Veness}\ \emph {et~al.}(2017)\citenamefont {Veness},
  \citenamefont {Essler},\ and\ \citenamefont {Fisher}}]{Veness2017}%
  \BibitemOpen
  \bibfield  {author} {\bibinfo {author} {\bibfnamefont {T.}~\bibnamefont
  {Veness}}, \bibinfo {author} {\bibfnamefont {F.~H.~L.}\ \bibnamefont
  {Essler}}, \ and\ \bibinfo {author} {\bibfnamefont {M.~P.~A.}\ \bibnamefont
  {Fisher}},\ }\href {\doibase 10.1103/PhysRevB.96.195153} {\bibfield
  {journal} {\bibinfo  {journal} {Phys. Rev. B}\ }\textbf {\bibinfo {volume}
  {96}},\ \bibinfo {pages} {195153} (\bibinfo {year} {2017})}\BibitemShut
  {NoStop}%
\bibitem [{\citenamefont {Yu}\ \emph {et~al.}(2018)\citenamefont {Yu},
  \citenamefont {Luo},\ and\ \citenamefont {Clark}}]{Clark2018}%
  \BibitemOpen
  \bibfield  {author} {\bibinfo {author} {\bibfnamefont {X.}~\bibnamefont
  {Yu}}, \bibinfo {author} {\bibfnamefont {D.}~\bibnamefont {Luo}}, \ and\
  \bibinfo {author} {\bibfnamefont {B.~K.}\ \bibnamefont {Clark}},\ }\href
  {\doibase 10.1103/PhysRevB.98.115106} {\bibfield  {journal} {\bibinfo
  {journal} {Phys. Rev. B}\ }\textbf {\bibinfo {volume} {98}},\ \bibinfo
  {pages} {115106} (\bibinfo {year} {2018})}\BibitemShut {NoStop}%
\bibitem [{\citenamefont {Moudgalya}\ \emph {et~al.}(2020)\citenamefont
  {Moudgalya}, \citenamefont {Regnault},\ and\ \citenamefont
  {Bernevig}}]{Bernevig2020}%
  \BibitemOpen
  \bibfield  {author} {\bibinfo {author} {\bibfnamefont {S.}~\bibnamefont
  {Moudgalya}}, \bibinfo {author} {\bibfnamefont {N.}~\bibnamefont {Regnault}},
  \ and\ \bibinfo {author} {\bibfnamefont {B.~A.}\ \bibnamefont {Bernevig}},\
  }\href {\doibase 10.1103/PhysRevB.102.085140} {\bibfield  {journal} {\bibinfo
   {journal} {Phys. Rev. B}\ }\textbf {\bibinfo {volume} {102}},\ \bibinfo
  {pages} {085140} (\bibinfo {year} {2020})}\BibitemShut {NoStop}%
\bibitem [{\citenamefont {Mark}\ and\ \citenamefont
  {Motrunich}(2020)}]{Motrunich2020}%
  \BibitemOpen
  \bibfield  {author} {\bibinfo {author} {\bibfnamefont {D.~K.}\ \bibnamefont
  {Mark}}\ and\ \bibinfo {author} {\bibfnamefont {O.~I.}\ \bibnamefont
  {Motrunich}},\ }\href {\doibase 10.1103/PhysRevB.102.075132} {\bibfield
  {journal} {\bibinfo  {journal} {Phys. Rev. B}\ }\textbf {\bibinfo {volume}
  {102}},\ \bibinfo {pages} {075132} (\bibinfo {year} {2020})}\BibitemShut
  {NoStop}%
\bibitem [{\citenamefont {Pesin}\ and\ \citenamefont {Balents}(2010)}]{PB2010}%
  \BibitemOpen
  \bibfield  {author} {\bibinfo {author} {\bibfnamefont {D.}~\bibnamefont
  {Pesin}}\ and\ \bibinfo {author} {\bibfnamefont {L.}~\bibnamefont
  {Balents}},\ }\href
  {http://www.nature.com/nphys/journal/v6/n5/full/nphys1606.html} {\bibfield
  {journal} {\bibinfo  {journal} {Nat. Phys.}\ }\textbf {\bibinfo {volume}
  {6}},\ \bibinfo {pages} {376} (\bibinfo {year} {2010})}\BibitemShut {NoStop}%
\bibitem [{\citenamefont {Witczak-Krempa}\ \emph
  {et~al.}(2014{\natexlab{a}})\citenamefont {Witczak-Krempa}, \citenamefont
  {Chen}, \citenamefont {Kim},\ and\ \citenamefont {Balents}}]{WCKB2014}%
  \BibitemOpen
  \bibfield  {author} {\bibinfo {author} {\bibfnamefont {W.}~\bibnamefont
  {Witczak-Krempa}}, \bibinfo {author} {\bibfnamefont {G.}~\bibnamefont
  {Chen}}, \bibinfo {author} {\bibfnamefont {Y.~B.}\ \bibnamefont {Kim}}, \
  and\ \bibinfo {author} {\bibfnamefont {L.}~\bibnamefont {Balents}},\ }\href
  {\doibase 10.1146/annurev-conmatphys-020911-125138} {\bibfield  {journal}
  {\bibinfo  {journal} {Annu. Rev. Condens. Matter Phys.}\ }\textbf {\bibinfo
  {volume} {5}},\ \bibinfo {pages} {57} (\bibinfo {year}
  {2014}{\natexlab{a}})}\BibitemShut {NoStop}%
\bibitem [{\citenamefont {Rau}\ \emph {et~al.}(2016)\citenamefont {Rau},
  \citenamefont {Lee},\ and\ \citenamefont {Kee}}]{Rau2016}%
  \BibitemOpen
  \bibfield  {author} {\bibinfo {author} {\bibfnamefont {J.~G.}\ \bibnamefont
  {Rau}}, \bibinfo {author} {\bibfnamefont {E.~K.-H.}\ \bibnamefont {Lee}}, \
  and\ \bibinfo {author} {\bibfnamefont {H.-Y.}\ \bibnamefont {Kee}},\ }\href
  {\doibase 10.1146/annurev-conmatphys-031115-011319} {\bibfield  {journal}
  {\bibinfo  {journal} {Annu. Rev. Condens. Matter Phys.}\ }\textbf {\bibinfo
  {volume} {7}},\ \bibinfo {pages} {null} (\bibinfo {year} {2016})}\BibitemShut
  {NoStop}%
\bibitem [{\citenamefont {Roberts}(1966{\natexlab{a}})}]{roberts1966a}%
  \BibitemOpen
  \bibfield  {author} {\bibinfo {author} {\bibfnamefont {J.~E.}\ \bibnamefont
  {Roberts}},\ }\href {\doibase 10.1063/1.1705001} {\bibfield  {journal}
  {\bibinfo  {journal} {Journal of Mathematical Physics}\ }\textbf {\bibinfo
  {volume} {7}},\ \bibinfo {pages} {1097} (\bibinfo {year}
  {1966}{\natexlab{a}})}\BibitemShut {NoStop}%
\bibitem [{\citenamefont {Roberts}(1966{\natexlab{b}})}]{roberts1966b}%
  \BibitemOpen
  \bibfield  {author} {\bibinfo {author} {\bibfnamefont {J.~E.}\ \bibnamefont
  {Roberts}},\ }\href {https://projecteuclid.org:443/euclid.cmp/1103839388}
  {\bibfield  {journal} {\bibinfo  {journal} {Comm. Math. Phys.}\ }\textbf
  {\bibinfo {volume} {3}},\ \bibinfo {pages} {98} (\bibinfo {year}
  {1966}{\natexlab{b}})}\BibitemShut {NoStop}%
\bibitem [{\citenamefont {Affleck}\ and\ \citenamefont
  {Marston}(1988)}]{Affleck1988}%
  \BibitemOpen
  \bibfield  {author} {\bibinfo {author} {\bibfnamefont {I.}~\bibnamefont
  {Affleck}}\ and\ \bibinfo {author} {\bibfnamefont {J.~B.}\ \bibnamefont
  {Marston}},\ }\href {\doibase 10.1103/PhysRevB.37.3774} {\bibfield  {journal}
  {\bibinfo  {journal} {Phys. Rev. B}\ }\textbf {\bibinfo {volume} {37}},\
  \bibinfo {pages} {3774} (\bibinfo {year} {1988})}\BibitemShut {NoStop}%
\bibitem [{\citenamefont {Dagotto}\ \emph {et~al.}(1988)\citenamefont
  {Dagotto}, \citenamefont {Fradkin},\ and\ \citenamefont
  {Moreo}}]{Dagotto1988}%
  \BibitemOpen
  \bibfield  {author} {\bibinfo {author} {\bibfnamefont {E.}~\bibnamefont
  {Dagotto}}, \bibinfo {author} {\bibfnamefont {E.}~\bibnamefont {Fradkin}}, \
  and\ \bibinfo {author} {\bibfnamefont {A.}~\bibnamefont {Moreo}},\ }\href
  {\doibase 10.1103/PhysRevB.38.2926} {\bibfield  {journal} {\bibinfo
  {journal} {Phys. Rev. B}\ }\textbf {\bibinfo {volume} {38}},\ \bibinfo
  {pages} {2926} (\bibinfo {year} {1988})}\BibitemShut {NoStop}%
\bibitem [{\citenamefont {Wen}(2002)}]{Wen2002}%
  \BibitemOpen
  \bibfield  {author} {\bibinfo {author} {\bibfnamefont {X.-G.}\ \bibnamefont
  {Wen}},\ }\href {\doibase 10.1103/PhysRevB.65.165113} {\bibfield  {journal}
  {\bibinfo  {journal} {Phys. Rev. B}\ }\textbf {\bibinfo {volume} {65}},\
  \bibinfo {pages} {165113} (\bibinfo {year} {2002})}\BibitemShut {NoStop}%
\bibitem [{\citenamefont {Jackiw}(1984)}]{Jackiw1984}%
  \BibitemOpen
  \bibfield  {author} {\bibinfo {author} {\bibfnamefont {R.}~\bibnamefont
  {Jackiw}},\ }\href {\doibase 10.1103/PhysRevD.29.2375} {\bibfield  {journal}
  {\bibinfo  {journal} {Phys. Rev. D}\ }\textbf {\bibinfo {volume} {29}},\
  \bibinfo {pages} {2375} (\bibinfo {year} {1984})}\BibitemShut {NoStop}%
\bibitem [{\citenamefont {Haldane}(1988)}]{Haldane1988}%
  \BibitemOpen
  \bibfield  {author} {\bibinfo {author} {\bibfnamefont {F.~D.~M.}\
  \bibnamefont {Haldane}},\ }\href {\doibase 10.1103/PhysRevLett.61.2015}
  {\bibfield  {journal} {\bibinfo  {journal} {Phys. Rev. Lett.}\ }\textbf
  {\bibinfo {volume} {61}},\ \bibinfo {pages} {2015} (\bibinfo {year}
  {1988})}\BibitemShut {NoStop}%
\bibitem [{\citenamefont {Melikyan}\ and\ \citenamefont {Te\ifmmode
  \check{s}\else \v{s}\fi{}anovi\ifmmode~\acute{c}\else
  \'{c}\fi{}}(2007)}]{Melikyan2007}%
  \BibitemOpen
  \bibfield  {author} {\bibinfo {author} {\bibfnamefont {A.}~\bibnamefont
  {Melikyan}}\ and\ \bibinfo {author} {\bibfnamefont {Z.}~\bibnamefont
  {Te\ifmmode \check{s}\else \v{s}\fi{}anovi\ifmmode~\acute{c}\else
  \'{c}\fi{}}},\ }\href {\doibase 10.1103/PhysRevB.76.094509} {\bibfield
  {journal} {\bibinfo  {journal} {Phys. Rev. B}\ }\textbf {\bibinfo {volume}
  {76}},\ \bibinfo {pages} {094509} (\bibinfo {year} {2007})}\BibitemShut
  {NoStop}%
\bibitem [{\citenamefont {Witczak-Krempa}\ \emph
  {et~al.}(2014{\natexlab{b}})\citenamefont {Witczak-Krempa}, \citenamefont
  {Knap},\ and\ \citenamefont {Abanin}}]{William2014}%
  \BibitemOpen
  \bibfield  {author} {\bibinfo {author} {\bibfnamefont {W.}~\bibnamefont
  {Witczak-Krempa}}, \bibinfo {author} {\bibfnamefont {M.}~\bibnamefont
  {Knap}}, \ and\ \bibinfo {author} {\bibfnamefont {D.}~\bibnamefont
  {Abanin}},\ }\href {\doibase 10.1103/PhysRevLett.113.136402} {\bibfield
  {journal} {\bibinfo  {journal} {Phys. Rev. Lett.}\ }\textbf {\bibinfo
  {volume} {113}},\ \bibinfo {pages} {136402} (\bibinfo {year}
  {2014}{\natexlab{b}})}\BibitemShut {NoStop}%
\bibitem [{\citenamefont {Roy}(2017)}]{Roy2017}%
  \BibitemOpen
  \bibfield  {author} {\bibinfo {author} {\bibfnamefont {B.}~\bibnamefont
  {Roy}},\ }\href {\doibase 10.1103/PhysRevB.96.041113} {\bibfield  {journal}
  {\bibinfo  {journal} {Phys. Rev. B}\ }\textbf {\bibinfo {volume} {96}},\
  \bibinfo {pages} {041113} (\bibinfo {year} {2017})}\BibitemShut {NoStop}%
\bibitem [{\citenamefont {Li}\ \emph {et~al.}(2016)\citenamefont {Li},
  \citenamefont {Yu}, \citenamefont {Gu},\ and\ \citenamefont {Li}}]{Kai2016}%
  \BibitemOpen
  \bibfield  {author} {\bibinfo {author} {\bibfnamefont {K.}~\bibnamefont
  {Li}}, \bibinfo {author} {\bibfnamefont {S.-L.}\ \bibnamefont {Yu}}, \bibinfo
  {author} {\bibfnamefont {Z.-L.}\ \bibnamefont {Gu}}, \ and\ \bibinfo {author}
  {\bibfnamefont {J.-X.}\ \bibnamefont {Li}},\ }\href {\doibase
  10.1103/PhysRevB.94.125120} {\bibfield  {journal} {\bibinfo  {journal} {Phys.
  Rev. B}\ }\textbf {\bibinfo {volume} {94}},\ \bibinfo {pages} {125120}
  (\bibinfo {year} {2016})}\BibitemShut {NoStop}%
\bibitem [{\citenamefont {Li}\ \emph {et~al.}(2015)\citenamefont {Li},
  \citenamefont {Yu},\ and\ \citenamefont {Li}}]{Kai2015}%
  \BibitemOpen
  \bibfield  {author} {\bibinfo {author} {\bibfnamefont {K.}~\bibnamefont
  {Li}}, \bibinfo {author} {\bibfnamefont {S.-L.}\ \bibnamefont {Yu}}, \ and\
  \bibinfo {author} {\bibfnamefont {J.-X.}\ \bibnamefont {Li}},\ }\href
  {http://stacks.iop.org/1367-2630/17/i=4/a=043032} {\bibfield  {journal}
  {\bibinfo  {journal} {New J. Phys.}\ }\textbf {\bibinfo {volume} {17}},\
  \bibinfo {pages} {043032} (\bibinfo {year} {2015})}\BibitemShut {NoStop}%
\bibitem [{\citenamefont {Zhai}(2005)}]{huizhai}%
  \BibitemOpen
  \bibfield  {author} {\bibinfo {author} {\bibfnamefont {H.}~\bibnamefont
  {Zhai}},\ }\href {\doibase 10.1103/PhysRevB.71.012512} {\bibfield  {journal}
  {\bibinfo  {journal} {Phys. Rev. B}\ }\textbf {\bibinfo {volume} {71}},\
  \bibinfo {pages} {012512} (\bibinfo {year} {2005})}\BibitemShut {NoStop}%
\bibitem [{\citenamefont {Pu}\ and\ \citenamefont {Shen}(1994)}]{Shen1994}%
  \BibitemOpen
  \bibfield  {author} {\bibinfo {author} {\bibfnamefont {F.-C.}\ \bibnamefont
  {Pu}}\ and\ \bibinfo {author} {\bibfnamefont {S.-Q.}\ \bibnamefont {Shen}},\
  }\href {\doibase 10.1103/PhysRevB.50.16086} {\bibfield  {journal} {\bibinfo
  {journal} {Phys. Rev. B}\ }\textbf {\bibinfo {volume} {50}},\ \bibinfo
  {pages} {16086} (\bibinfo {year} {1994})}\BibitemShut {NoStop}%
\bibitem [{\citenamefont {Fujiuchi}\ \emph {et~al.}(2019)\citenamefont
  {Fujiuchi}, \citenamefont {Kaneko}, \citenamefont {Ohta},\ and\ \citenamefont
  {Yunoki}}]{Yunoki2019}%
  \BibitemOpen
  \bibfield  {author} {\bibinfo {author} {\bibfnamefont {R.}~\bibnamefont
  {Fujiuchi}}, \bibinfo {author} {\bibfnamefont {T.}~\bibnamefont {Kaneko}},
  \bibinfo {author} {\bibfnamefont {Y.}~\bibnamefont {Ohta}}, \ and\ \bibinfo
  {author} {\bibfnamefont {S.}~\bibnamefont {Yunoki}},\ }\href {\doibase
  10.1103/PhysRevB.100.045121} {\bibfield  {journal} {\bibinfo  {journal}
  {Phys. Rev. B}\ }\textbf {\bibinfo {volume} {100}},\ \bibinfo {pages}
  {045121} (\bibinfo {year} {2019})}\BibitemShut {NoStop}%
\bibitem [{\citenamefont {Fava}\ \emph {et~al.}(2020)\citenamefont {Fava},
  \citenamefont {Ware}, \citenamefont {Gopalakrishnan}, \citenamefont
  {Vasseur},\ and\ \citenamefont {Parameswaran}}]{Parameswaran2020}%
  \BibitemOpen
  \bibfield  {author} {\bibinfo {author} {\bibfnamefont {M.}~\bibnamefont
  {Fava}}, \bibinfo {author} {\bibfnamefont {B.}~\bibnamefont {Ware}}, \bibinfo
  {author} {\bibfnamefont {S.}~\bibnamefont {Gopalakrishnan}}, \bibinfo
  {author} {\bibfnamefont {R.}~\bibnamefont {Vasseur}}, \ and\ \bibinfo
  {author} {\bibfnamefont {S.~A.}\ \bibnamefont {Parameswaran}},\ }\href
  {\doibase 10.1103/PhysRevB.102.115121} {\bibfield  {journal} {\bibinfo
  {journal} {Phys. Rev. B}\ }\textbf {\bibinfo {volume} {102}},\ \bibinfo
  {pages} {115121} (\bibinfo {year} {2020})}\BibitemShut {NoStop}%
\bibitem [{\citenamefont {de~Boer}\ and\ \citenamefont
  {Schadschneider}(1995)}]{Boer2}%
  \BibitemOpen
  \bibfield  {author} {\bibinfo {author} {\bibfnamefont {J.}~\bibnamefont
  {de~Boer}}\ and\ \bibinfo {author} {\bibfnamefont {A.}~\bibnamefont
  {Schadschneider}},\ }\href {\doibase 10.1103/PhysRevLett.75.4298} {\bibfield
  {journal} {\bibinfo  {journal} {Phys. Rev. Lett.}\ }\textbf {\bibinfo
  {volume} {75}},\ \bibinfo {pages} {4298} (\bibinfo {year}
  {1995})}\BibitemShut {NoStop}%
\bibitem [{\citenamefont {Kivelson}\ \emph {et~al.}(1987)\citenamefont
  {Kivelson}, \citenamefont {Su}, \citenamefont {Schrieffer},\ and\
  \citenamefont {Heeger}}]{KSSH}%
  \BibitemOpen
  \bibfield  {author} {\bibinfo {author} {\bibfnamefont {S.}~\bibnamefont
  {Kivelson}}, \bibinfo {author} {\bibfnamefont {W.-P.}\ \bibnamefont {Su}},
  \bibinfo {author} {\bibfnamefont {J.~R.}\ \bibnamefont {Schrieffer}}, \ and\
  \bibinfo {author} {\bibfnamefont {A.~J.}\ \bibnamefont {Heeger}},\ }\href
  {\doibase 10.1103/PhysRevLett.58.1899} {\bibfield  {journal} {\bibinfo
  {journal} {Phys. Rev. Lett.}\ }\textbf {\bibinfo {volume} {58}},\ \bibinfo
  {pages} {1899} (\bibinfo {year} {1987})}\BibitemShut {NoStop}%
\bibitem [{\citenamefont {Montorsi}\ and\ \citenamefont
  {Rasetti}(1991)}]{Montorsi1991}%
  \BibitemOpen
  \bibfield  {author} {\bibinfo {author} {\bibfnamefont {A.}~\bibnamefont
  {Montorsi}}\ and\ \bibinfo {author} {\bibfnamefont {M.}~\bibnamefont
  {Rasetti}},\ }\href {\doibase 10.1103/PhysRevLett.66.1383} {\bibfield
  {journal} {\bibinfo  {journal} {Phys. Rev. Lett.}\ }\textbf {\bibinfo
  {volume} {66}},\ \bibinfo {pages} {1383} (\bibinfo {year}
  {1991})}\BibitemShut {NoStop}%
\bibitem [{\citenamefont {Yang}(1962)}]{Yang1962}%
  \BibitemOpen
  \bibfield  {author} {\bibinfo {author} {\bibfnamefont {C.~N.}\ \bibnamefont
  {Yang}},\ }\href {\doibase 10.1103/RevModPhys.34.694} {\bibfield  {journal}
  {\bibinfo  {journal} {Rev. Mod. Phys.}\ }\textbf {\bibinfo {volume} {34}},\
  \bibinfo {pages} {694} (\bibinfo {year} {1962})}\BibitemShut {NoStop}%
\bibitem [{\citenamefont {Sehayek}\ \emph {et~al.}(2020)\citenamefont
  {Sehayek}, \citenamefont {Thakurathi},\ and\ \citenamefont
  {Burkov}}]{Burkov2020}%
  \BibitemOpen
  \bibfield  {author} {\bibinfo {author} {\bibfnamefont {D.}~\bibnamefont
  {Sehayek}}, \bibinfo {author} {\bibfnamefont {M.}~\bibnamefont {Thakurathi}},
  \ and\ \bibinfo {author} {\bibfnamefont {A.}~\bibnamefont {Burkov}},\ }\href
  {https://arxiv.org/abs/2007.07256} {\bibfield  {journal} {\bibinfo  {journal}
  {arXiv:2007.07256}\ } (\bibinfo {year} {2020})}\BibitemShut {NoStop}%
\bibitem [{\citenamefont {Dalibard}\ \emph {et~al.}(2011)\citenamefont
  {Dalibard}, \citenamefont {Gerbier}, \citenamefont
  {Juzeli\ifmmode~\bar{u}\else \={u}\fi{}nas},\ and\ \citenamefont
  {\"Ohberg}}]{Dalibard2011}%
  \BibitemOpen
  \bibfield  {author} {\bibinfo {author} {\bibfnamefont {J.}~\bibnamefont
  {Dalibard}}, \bibinfo {author} {\bibfnamefont {F.}~\bibnamefont {Gerbier}},
  \bibinfo {author} {\bibfnamefont {G.}~\bibnamefont
  {Juzeli\ifmmode~\bar{u}\else \={u}\fi{}nas}}, \ and\ \bibinfo {author}
  {\bibfnamefont {P.}~\bibnamefont {\"Ohberg}},\ }\href {\doibase
  10.1103/RevModPhys.83.1523} {\bibfield  {journal} {\bibinfo  {journal} {Rev.
  Mod. Phys.}\ }\textbf {\bibinfo {volume} {83}},\ \bibinfo {pages} {1523}
  (\bibinfo {year} {2011})}\BibitemShut {NoStop}%
\bibitem [{\citenamefont {Chin}\ \emph {et~al.}(2010)\citenamefont {Chin},
  \citenamefont {Grimm}, \citenamefont {Julienne},\ and\ \citenamefont
  {Tiesinga}}]{Chinrmp2010}%
  \BibitemOpen
  \bibfield  {author} {\bibinfo {author} {\bibfnamefont {C.}~\bibnamefont
  {Chin}}, \bibinfo {author} {\bibfnamefont {R.}~\bibnamefont {Grimm}},
  \bibinfo {author} {\bibfnamefont {P.}~\bibnamefont {Julienne}}, \ and\
  \bibinfo {author} {\bibfnamefont {E.}~\bibnamefont {Tiesinga}},\ }\href
  {\doibase 10.1103/RevModPhys.82.1225} {\bibfield  {journal} {\bibinfo
  {journal} {Rev. Mod. Phys.}\ }\textbf {\bibinfo {volume} {82}},\ \bibinfo
  {pages} {1225} (\bibinfo {year} {2010})}\BibitemShut {NoStop}%
\bibitem [{\citenamefont {Giorgini}\ \emph {et~al.}(2008)\citenamefont
  {Giorgini}, \citenamefont {Pitaevskii},\ and\ \citenamefont
  {Stringari}}]{Giorginirmp2008}%
  \BibitemOpen
  \bibfield  {author} {\bibinfo {author} {\bibfnamefont {S.}~\bibnamefont
  {Giorgini}}, \bibinfo {author} {\bibfnamefont {L.~P.}\ \bibnamefont
  {Pitaevskii}}, \ and\ \bibinfo {author} {\bibfnamefont {S.}~\bibnamefont
  {Stringari}},\ }\href {\doibase 10.1103/RevModPhys.80.1215} {\bibfield
  {journal} {\bibinfo  {journal} {Rev. Mod. Phys.}\ }\textbf {\bibinfo {volume}
  {80}},\ \bibinfo {pages} {1215} (\bibinfo {year} {2008})}\BibitemShut
  {NoStop}%
\bibitem [{\citenamefont {Duan}\ \emph {et~al.}(2003)\citenamefont {Duan},
  \citenamefont {Demler},\ and\ \citenamefont {Lukin}}]{Duan2003}%
  \BibitemOpen
  \bibfield  {author} {\bibinfo {author} {\bibfnamefont {L.-M.}\ \bibnamefont
  {Duan}}, \bibinfo {author} {\bibfnamefont {E.}~\bibnamefont {Demler}}, \ and\
  \bibinfo {author} {\bibfnamefont {M.~D.}\ \bibnamefont {Lukin}},\ }\href
  {\doibase 10.1103/PhysRevLett.91.090402} {\bibfield  {journal} {\bibinfo
  {journal} {Phys. Rev. Lett.}\ }\textbf {\bibinfo {volume} {91}},\ \bibinfo
  {pages} {090402} (\bibinfo {year} {2003})}\BibitemShut {NoStop}%
\bibitem [{\citenamefont {Liu}\ \emph {et~al.}(2004)\citenamefont {Liu},
  \citenamefont {Wilczek},\ and\ \citenamefont {Zoller}}]{Zoller2004}%
  \BibitemOpen
  \bibfield  {author} {\bibinfo {author} {\bibfnamefont {W.~V.}\ \bibnamefont
  {Liu}}, \bibinfo {author} {\bibfnamefont {F.}~\bibnamefont {Wilczek}}, \ and\
  \bibinfo {author} {\bibfnamefont {P.}~\bibnamefont {Zoller}},\ }\href
  {\doibase 10.1103/PhysRevA.70.033603} {\bibfield  {journal} {\bibinfo
  {journal} {Phys. Rev. A}\ }\textbf {\bibinfo {volume} {70}},\ \bibinfo
  {pages} {033603} (\bibinfo {year} {2004})}\BibitemShut {NoStop}%
\bibitem [{\citenamefont {Schneider}\ \emph {et~al.}(2008)\citenamefont
  {Schneider}, \citenamefont {Hackerm{\"u}ller}, \citenamefont {Will},
  \citenamefont {Best}, \citenamefont {Bloch}, \citenamefont {Costi},
  \citenamefont {Helmes}, \citenamefont {Rasch},\ and\ \citenamefont
  {Rosch}}]{Schneider2008}%
  \BibitemOpen
  \bibfield  {author} {\bibinfo {author} {\bibfnamefont {U.}~\bibnamefont
  {Schneider}}, \bibinfo {author} {\bibfnamefont {L.}~\bibnamefont
  {Hackerm{\"u}ller}}, \bibinfo {author} {\bibfnamefont {S.}~\bibnamefont
  {Will}}, \bibinfo {author} {\bibfnamefont {T.}~\bibnamefont {Best}}, \bibinfo
  {author} {\bibfnamefont {I.}~\bibnamefont {Bloch}}, \bibinfo {author}
  {\bibfnamefont {T.~A.}\ \bibnamefont {Costi}}, \bibinfo {author}
  {\bibfnamefont {R.~W.}\ \bibnamefont {Helmes}}, \bibinfo {author}
  {\bibfnamefont {D.}~\bibnamefont {Rasch}}, \ and\ \bibinfo {author}
  {\bibfnamefont {A.}~\bibnamefont {Rosch}},\ }\href {\doibase
  10.1126/science.1165449} {\bibfield  {journal} {\bibinfo  {journal}
  {Science}\ }\textbf {\bibinfo {volume} {322}},\ \bibinfo {pages} {1520}
  (\bibinfo {year} {2008})}\BibitemShut {NoStop}%
\bibitem [{\citenamefont {Jordens}\ \emph {et~al.}(2008)\citenamefont
  {Jordens}, \citenamefont {Strohmaier}, \citenamefont {Gunter}, \citenamefont
  {Moritz},\ and\ \citenamefont {Esslinger}}]{Jordens2008}%
  \BibitemOpen
  \bibfield  {author} {\bibinfo {author} {\bibfnamefont {R.}~\bibnamefont
  {Jordens}}, \bibinfo {author} {\bibfnamefont {N.}~\bibnamefont {Strohmaier}},
  \bibinfo {author} {\bibfnamefont {K.}~\bibnamefont {Gunter}}, \bibinfo
  {author} {\bibfnamefont {H.}~\bibnamefont {Moritz}}, \ and\ \bibinfo {author}
  {\bibfnamefont {T.}~\bibnamefont {Esslinger}},\ }\href {\doibase
  10.1038/nature07244} {\bibfield  {journal} {\bibinfo  {journal} {Nature}\
  }\textbf {\bibinfo {volume} {455}},\ \bibinfo {pages} {204} (\bibinfo {year}
  {2008})}\BibitemShut {NoStop}%
\bibitem [{\citenamefont {Strohmaier}\ \emph {et~al.}(2007)\citenamefont
  {Strohmaier}, \citenamefont {Takasu}, \citenamefont {G\"unter}, \citenamefont
  {J\"ordens}, \citenamefont {K\"ohl}, \citenamefont {Moritz},\ and\
  \citenamefont {Esslinger}}]{Strohmaier2007}%
  \BibitemOpen
  \bibfield  {author} {\bibinfo {author} {\bibfnamefont {N.}~\bibnamefont
  {Strohmaier}}, \bibinfo {author} {\bibfnamefont {Y.}~\bibnamefont {Takasu}},
  \bibinfo {author} {\bibfnamefont {K.}~\bibnamefont {G\"unter}}, \bibinfo
  {author} {\bibfnamefont {R.}~\bibnamefont {J\"ordens}}, \bibinfo {author}
  {\bibfnamefont {M.}~\bibnamefont {K\"ohl}}, \bibinfo {author} {\bibfnamefont
  {H.}~\bibnamefont {Moritz}}, \ and\ \bibinfo {author} {\bibfnamefont
  {T.}~\bibnamefont {Esslinger}},\ }\href {\doibase
  10.1103/PhysRevLett.99.220601} {\bibfield  {journal} {\bibinfo  {journal}
  {Phys. Rev. Lett.}\ }\textbf {\bibinfo {volume} {99}},\ \bibinfo {pages}
  {220601} (\bibinfo {year} {2007})}\BibitemShut {NoStop}%
\bibitem [{\citenamefont {Hart}\ \emph {et~al.}(2015)\citenamefont {Hart},
  \citenamefont {Duarte}, \citenamefont {Yang}, \citenamefont {Liu},
  \citenamefont {Paiva}, \citenamefont {Khatami}, \citenamefont {Scalettar},
  \citenamefont {Trivedi}, \citenamefont {Huse},\ and\ \citenamefont
  {Hulet}}]{Hart2015}%
  \BibitemOpen
  \bibfield  {author} {\bibinfo {author} {\bibfnamefont {R.~A.}\ \bibnamefont
  {Hart}}, \bibinfo {author} {\bibfnamefont {P.~M.}\ \bibnamefont {Duarte}},
  \bibinfo {author} {\bibfnamefont {T.-L.}\ \bibnamefont {Yang}}, \bibinfo
  {author} {\bibfnamefont {X.}~\bibnamefont {Liu}}, \bibinfo {author}
  {\bibfnamefont {T.}~\bibnamefont {Paiva}}, \bibinfo {author} {\bibfnamefont
  {E.}~\bibnamefont {Khatami}}, \bibinfo {author} {\bibfnamefont {R.~T.}\
  \bibnamefont {Scalettar}}, \bibinfo {author} {\bibfnamefont {N.}~\bibnamefont
  {Trivedi}}, \bibinfo {author} {\bibfnamefont {D.~A.}\ \bibnamefont {Huse}}, \
  and\ \bibinfo {author} {\bibfnamefont {R.~G.}\ \bibnamefont {Hulet}},\ }\href
  {\doibase 10.1038/nature14223} {\bibfield  {journal} {\bibinfo  {journal}
  {Nature}\ }\textbf {\bibinfo {volume} {519}},\ \bibinfo {pages} {211}
  (\bibinfo {year} {2015})}\BibitemShut {NoStop}%
\bibitem [{\citenamefont {Mazurenko}\ \emph {et~al.}(2017)\citenamefont
  {Mazurenko}, \citenamefont {Chiu}, \citenamefont {Ji}, \citenamefont
  {Parsons}, \citenamefont {Kanasz-Nagy}, \citenamefont {Schmidt},
  \citenamefont {Grusdt}, \citenamefont {Demler}, \citenamefont {Greif},\ and\
  \citenamefont {Greiner}}]{Mazurenko2017}%
  \BibitemOpen
  \bibfield  {author} {\bibinfo {author} {\bibfnamefont {A.}~\bibnamefont
  {Mazurenko}}, \bibinfo {author} {\bibfnamefont {C.~S.}\ \bibnamefont {Chiu}},
  \bibinfo {author} {\bibfnamefont {G.}~\bibnamefont {Ji}}, \bibinfo {author}
  {\bibfnamefont {M.~F.}\ \bibnamefont {Parsons}}, \bibinfo {author}
  {\bibfnamefont {M.}~\bibnamefont {Kanasz-Nagy}}, \bibinfo {author}
  {\bibfnamefont {R.}~\bibnamefont {Schmidt}}, \bibinfo {author} {\bibfnamefont
  {F.}~\bibnamefont {Grusdt}}, \bibinfo {author} {\bibfnamefont
  {E.}~\bibnamefont {Demler}}, \bibinfo {author} {\bibfnamefont
  {D.}~\bibnamefont {Greif}}, \ and\ \bibinfo {author} {\bibfnamefont
  {M.}~\bibnamefont {Greiner}},\ }\href {\doibase 10.1038/nature22362}
  {\bibfield  {journal} {\bibinfo  {journal} {Nature}\ }\textbf {\bibinfo
  {volume} {545}},\ \bibinfo {pages} {462} (\bibinfo {year}
  {2017})}\BibitemShut {NoStop}%
\bibitem [{\citenamefont {Parsons}\ \emph {et~al.}(2016)\citenamefont
  {Parsons}, \citenamefont {Mazurenko}, \citenamefont {Chiu}, \citenamefont
  {Ji}, \citenamefont {Greif},\ and\ \citenamefont {Greiner}}]{Parsons2016}%
  \BibitemOpen
  \bibfield  {author} {\bibinfo {author} {\bibfnamefont {M.~F.}\ \bibnamefont
  {Parsons}}, \bibinfo {author} {\bibfnamefont {A.}~\bibnamefont {Mazurenko}},
  \bibinfo {author} {\bibfnamefont {C.~S.}\ \bibnamefont {Chiu}}, \bibinfo
  {author} {\bibfnamefont {G.}~\bibnamefont {Ji}}, \bibinfo {author}
  {\bibfnamefont {D.}~\bibnamefont {Greif}}, \ and\ \bibinfo {author}
  {\bibfnamefont {M.}~\bibnamefont {Greiner}},\ }\href {\doibase
  10.1126/science.aag1430} {\bibfield  {journal} {\bibinfo  {journal}
  {Science}\ }\textbf {\bibinfo {volume} {353}},\ \bibinfo {pages} {1253}
  (\bibinfo {year} {2016})}\BibitemShut {NoStop}%
\bibitem [{\citenamefont {Boll}\ \emph {et~al.}(2016)\citenamefont {Boll},
  \citenamefont {Hilker}, \citenamefont {Salomon}, \citenamefont {Omran},
  \citenamefont {Nespolo}, \citenamefont {Pollet}, \citenamefont {Bloch},\ and\
  \citenamefont {Gross}}]{Boll2016}%
  \BibitemOpen
  \bibfield  {author} {\bibinfo {author} {\bibfnamefont {M.}~\bibnamefont
  {Boll}}, \bibinfo {author} {\bibfnamefont {T.~A.}\ \bibnamefont {Hilker}},
  \bibinfo {author} {\bibfnamefont {G.}~\bibnamefont {Salomon}}, \bibinfo
  {author} {\bibfnamefont {A.}~\bibnamefont {Omran}}, \bibinfo {author}
  {\bibfnamefont {J.}~\bibnamefont {Nespolo}}, \bibinfo {author} {\bibfnamefont
  {L.}~\bibnamefont {Pollet}}, \bibinfo {author} {\bibfnamefont
  {I.}~\bibnamefont {Bloch}}, \ and\ \bibinfo {author} {\bibfnamefont
  {C.}~\bibnamefont {Gross}},\ }\href {\doibase 10.1126/science.aag1635}
  {\bibfield  {journal} {\bibinfo  {journal} {Science}\ }\textbf {\bibinfo
  {volume} {353}},\ \bibinfo {pages} {1257} (\bibinfo {year}
  {2016})}\BibitemShut {NoStop}%
\bibitem [{\citenamefont {Cheuk}\ \emph {et~al.}(2016)\citenamefont {Cheuk},
  \citenamefont {Nichols}, \citenamefont {Lawrence}, \citenamefont {Okan},
  \citenamefont {Zhang}, \citenamefont {Khatami}, \citenamefont {Trivedi},
  \citenamefont {Paiva}, \citenamefont {Rigol},\ and\ \citenamefont
  {Zwierlein}}]{Cheuk2016}%
  \BibitemOpen
  \bibfield  {author} {\bibinfo {author} {\bibfnamefont {L.~W.}\ \bibnamefont
  {Cheuk}}, \bibinfo {author} {\bibfnamefont {M.~A.}\ \bibnamefont {Nichols}},
  \bibinfo {author} {\bibfnamefont {K.~R.}\ \bibnamefont {Lawrence}}, \bibinfo
  {author} {\bibfnamefont {M.}~\bibnamefont {Okan}}, \bibinfo {author}
  {\bibfnamefont {H.}~\bibnamefont {Zhang}}, \bibinfo {author} {\bibfnamefont
  {E.}~\bibnamefont {Khatami}}, \bibinfo {author} {\bibfnamefont
  {N.}~\bibnamefont {Trivedi}}, \bibinfo {author} {\bibfnamefont
  {T.}~\bibnamefont {Paiva}}, \bibinfo {author} {\bibfnamefont
  {M.}~\bibnamefont {Rigol}}, \ and\ \bibinfo {author} {\bibfnamefont {M.~W.}\
  \bibnamefont {Zwierlein}},\ }\href {\doibase 10.1126/science.aag3349}
  {\bibfield  {journal} {\bibinfo  {journal} {Science}\ }\textbf {\bibinfo
  {volume} {353}},\ \bibinfo {pages} {1260} (\bibinfo {year}
  {2016})}\BibitemShut {NoStop}%
\bibitem [{\citenamefont {Brown}\ \emph {et~al.}(2017)\citenamefont {Brown},
  \citenamefont {Mitra}, \citenamefont {Guardado-Sanchez}, \citenamefont
  {Schau{\ss}}, \citenamefont {Kondov}, \citenamefont {Khatami}, \citenamefont
  {Paiva}, \citenamefont {Trivedi}, \citenamefont {Huse},\ and\ \citenamefont
  {Bakr}}]{Brown2017}%
  \BibitemOpen
  \bibfield  {author} {\bibinfo {author} {\bibfnamefont {P.~T.}\ \bibnamefont
  {Brown}}, \bibinfo {author} {\bibfnamefont {D.}~\bibnamefont {Mitra}},
  \bibinfo {author} {\bibfnamefont {E.}~\bibnamefont {Guardado-Sanchez}},
  \bibinfo {author} {\bibfnamefont {P.}~\bibnamefont {Schau{\ss}}}, \bibinfo
  {author} {\bibfnamefont {S.~S.}\ \bibnamefont {Kondov}}, \bibinfo {author}
  {\bibfnamefont {E.}~\bibnamefont {Khatami}}, \bibinfo {author} {\bibfnamefont
  {T.}~\bibnamefont {Paiva}}, \bibinfo {author} {\bibfnamefont
  {N.}~\bibnamefont {Trivedi}}, \bibinfo {author} {\bibfnamefont {D.~A.}\
  \bibnamefont {Huse}}, \ and\ \bibinfo {author} {\bibfnamefont {W.~S.}\
  \bibnamefont {Bakr}},\ }\href {\doibase 10.1126/science.aam7838} {\bibfield
  {journal} {\bibinfo  {journal} {Science}\ }\textbf {\bibinfo {volume}
  {357}},\ \bibinfo {pages} {1385} (\bibinfo {year} {2017})}\BibitemShut
  {NoStop}%
\bibitem [{\citenamefont {Bloch}\ \emph {et~al.}(2008)\citenamefont {Bloch},
  \citenamefont {Dalibard},\ and\ \citenamefont {Zwerger}}]{Bloch2008}%
  \BibitemOpen
  \bibfield  {author} {\bibinfo {author} {\bibfnamefont {I.}~\bibnamefont
  {Bloch}}, \bibinfo {author} {\bibfnamefont {J.}~\bibnamefont {Dalibard}}, \
  and\ \bibinfo {author} {\bibfnamefont {W.}~\bibnamefont {Zwerger}},\ }\href
  {\doibase 10.1103/RevModPhys.80.885} {\bibfield  {journal} {\bibinfo
  {journal} {Rev. Mod. Phys.}\ }\textbf {\bibinfo {volume} {80}},\ \bibinfo
  {pages} {885} (\bibinfo {year} {2008})}\BibitemShut {NoStop}%
\bibitem [{\citenamefont {Werner}\ \emph {et~al.}(2005)\citenamefont {Werner},
  \citenamefont {Parcollet}, \citenamefont {Georges},\ and\ \citenamefont
  {Hassan}}]{Werner2005}%
  \BibitemOpen
  \bibfield  {author} {\bibinfo {author} {\bibfnamefont {F.}~\bibnamefont
  {Werner}}, \bibinfo {author} {\bibfnamefont {O.}~\bibnamefont {Parcollet}},
  \bibinfo {author} {\bibfnamefont {A.}~\bibnamefont {Georges}}, \ and\
  \bibinfo {author} {\bibfnamefont {S.~R.}\ \bibnamefont {Hassan}},\ }\href
  {\doibase 10.1103/PhysRevLett.95.056401} {\bibfield  {journal} {\bibinfo
  {journal} {Phys. Rev. Lett.}\ }\textbf {\bibinfo {volume} {95}},\ \bibinfo
  {pages} {056401} (\bibinfo {year} {2005})}\BibitemShut {NoStop}%
\bibitem [{\citenamefont {Rosch}\ \emph {et~al.}(2008)\citenamefont {Rosch},
  \citenamefont {Rasch}, \citenamefont {Binz},\ and\ \citenamefont
  {Vojta}}]{Rosch2008}%
  \BibitemOpen
  \bibfield  {author} {\bibinfo {author} {\bibfnamefont {A.}~\bibnamefont
  {Rosch}}, \bibinfo {author} {\bibfnamefont {D.}~\bibnamefont {Rasch}},
  \bibinfo {author} {\bibfnamefont {B.}~\bibnamefont {Binz}}, \ and\ \bibinfo
  {author} {\bibfnamefont {M.}~\bibnamefont {Vojta}},\ }\href {\doibase
  10.1103/PhysRevLett.101.265301} {\bibfield  {journal} {\bibinfo  {journal}
  {Phys. Rev. Lett.}\ }\textbf {\bibinfo {volume} {101}},\ \bibinfo {pages}
  {265301} (\bibinfo {year} {2008})}\BibitemShut {NoStop}%
\bibitem [{\citenamefont {Li}\ and\ \citenamefont {Sarma}(2015)}]{Sarma2015}%
  \BibitemOpen
  \bibfield  {author} {\bibinfo {author} {\bibfnamefont {X.}~\bibnamefont
  {Li}}\ and\ \bibinfo {author} {\bibfnamefont {S.~D.}\ \bibnamefont {Sarma}},\
  }\href {\doibase 10.1038/ncomms8137} {\bibfield  {journal} {\bibinfo
  {journal} {Nature Communications}\ }\textbf {\bibinfo {volume} {6}},\
  \bibinfo {pages} {7137} (\bibinfo {year} {2015})}\BibitemShut {NoStop}%
\bibitem [{\citenamefont {Carroll}\ \emph {et~al.}(2004)\citenamefont
  {Carroll}, \citenamefont {Claringbould}, \citenamefont {Goodsell},
  \citenamefont {Lim},\ and\ \citenamefont {Noel}}]{Carroll2004}%
  \BibitemOpen
  \bibfield  {author} {\bibinfo {author} {\bibfnamefont {T.~J.}\ \bibnamefont
  {Carroll}}, \bibinfo {author} {\bibfnamefont {K.}~\bibnamefont
  {Claringbould}}, \bibinfo {author} {\bibfnamefont {A.}~\bibnamefont
  {Goodsell}}, \bibinfo {author} {\bibfnamefont {M.~J.}\ \bibnamefont {Lim}}, \
  and\ \bibinfo {author} {\bibfnamefont {M.~W.}\ \bibnamefont {Noel}},\ }\href
  {\doibase 10.1103/PhysRevLett.93.153001} {\bibfield  {journal} {\bibinfo
  {journal} {Phys. Rev. Lett.}\ }\textbf {\bibinfo {volume} {93}},\ \bibinfo
  {pages} {153001} (\bibinfo {year} {2004})}\BibitemShut {NoStop}%
\bibitem [{\citenamefont {Viteau}\ \emph {et~al.}(2011)\citenamefont {Viteau},
  \citenamefont {Bason}, \citenamefont {Radogostowicz}, \citenamefont
  {Malossi}, \citenamefont {Ciampini}, \citenamefont {Morsch},\ and\
  \citenamefont {Arimondo}}]{Viteau2011}%
  \BibitemOpen
  \bibfield  {author} {\bibinfo {author} {\bibfnamefont {M.}~\bibnamefont
  {Viteau}}, \bibinfo {author} {\bibfnamefont {M.~G.}\ \bibnamefont {Bason}},
  \bibinfo {author} {\bibfnamefont {J.}~\bibnamefont {Radogostowicz}}, \bibinfo
  {author} {\bibfnamefont {N.}~\bibnamefont {Malossi}}, \bibinfo {author}
  {\bibfnamefont {D.}~\bibnamefont {Ciampini}}, \bibinfo {author}
  {\bibfnamefont {O.}~\bibnamefont {Morsch}}, \ and\ \bibinfo {author}
  {\bibfnamefont {E.}~\bibnamefont {Arimondo}},\ }\href {\doibase
  10.1103/PhysRevLett.107.060402} {\bibfield  {journal} {\bibinfo  {journal}
  {Phys. Rev. Lett.}\ }\textbf {\bibinfo {volume} {107}},\ \bibinfo {pages}
  {060402} (\bibinfo {year} {2011})}\BibitemShut {NoStop}%
\bibitem [{\citenamefont {Schauss}\ \emph {et~al.}(2012)\citenamefont
  {Schauss}, \citenamefont {Cheneau}, \citenamefont {Endres}, \citenamefont
  {Fukuhara}, \citenamefont {Hild}, \citenamefont {Omran}, \citenamefont
  {Pohl}, \citenamefont {Gross}, \citenamefont {Kuhr},\ and\ \citenamefont
  {Bloch}}]{Bloch2012}%
  \BibitemOpen
  \bibfield  {author} {\bibinfo {author} {\bibfnamefont {P.}~\bibnamefont
  {Schauss}}, \bibinfo {author} {\bibfnamefont {M.}~\bibnamefont {Cheneau}},
  \bibinfo {author} {\bibfnamefont {M.}~\bibnamefont {Endres}}, \bibinfo
  {author} {\bibfnamefont {T.}~\bibnamefont {Fukuhara}}, \bibinfo {author}
  {\bibfnamefont {S.}~\bibnamefont {Hild}}, \bibinfo {author} {\bibfnamefont
  {A.}~\bibnamefont {Omran}}, \bibinfo {author} {\bibfnamefont
  {T.}~\bibnamefont {Pohl}}, \bibinfo {author} {\bibfnamefont {C.}~\bibnamefont
  {Gross}}, \bibinfo {author} {\bibfnamefont {S.}~\bibnamefont {Kuhr}}, \ and\
  \bibinfo {author} {\bibfnamefont {I.}~\bibnamefont {Bloch}},\ }\href
  {\doibase 10.1038/nature11596} {\bibfield  {journal} {\bibinfo  {journal}
  {Nature}\ }\textbf {\bibinfo {volume} {491}},\ \bibinfo {pages} {87}
  (\bibinfo {year} {2012})}\BibitemShut {NoStop}%
\bibitem [{\citenamefont {Labuhn}\ \emph {et~al.}(2016)\citenamefont {Labuhn},
  \citenamefont {Barredo}, \citenamefont {Ravets}, \citenamefont {de~Leseleuc},
  \citenamefont {Macri}, \citenamefont {Lahaye},\ and\ \citenamefont
  {Browaeys}}]{Labuhn2016}%
  \BibitemOpen
  \bibfield  {author} {\bibinfo {author} {\bibfnamefont {H.}~\bibnamefont
  {Labuhn}}, \bibinfo {author} {\bibfnamefont {D.}~\bibnamefont {Barredo}},
  \bibinfo {author} {\bibfnamefont {S.}~\bibnamefont {Ravets}}, \bibinfo
  {author} {\bibfnamefont {S.}~\bibnamefont {de~Leseleuc}}, \bibinfo {author}
  {\bibfnamefont {T.}~\bibnamefont {Macri}}, \bibinfo {author} {\bibfnamefont
  {T.}~\bibnamefont {Lahaye}}, \ and\ \bibinfo {author} {\bibfnamefont
  {A.}~\bibnamefont {Browaeys}},\ }\href {\doibase 10.1038/nature18274}
  {\bibfield  {journal} {\bibinfo  {journal} {Nature}\ }\textbf {\bibinfo
  {volume} {534}},\ \bibinfo {pages} {667} (\bibinfo {year}
  {2016})}\BibitemShut {NoStop}%
\bibitem [{\citenamefont {Chin}\ \emph {et~al.}(2006)\citenamefont {Chin},
  \citenamefont {Miller}, \citenamefont {Liu}, \citenamefont {Stan},
  \citenamefont {Setiawan}, \citenamefont {Sanner}, \citenamefont {Xu},\ and\
  \citenamefont {Ketterle}}]{Chin2006}%
  \BibitemOpen
  \bibfield  {author} {\bibinfo {author} {\bibfnamefont {J.~K.}\ \bibnamefont
  {Chin}}, \bibinfo {author} {\bibfnamefont {D.~E.}\ \bibnamefont {Miller}},
  \bibinfo {author} {\bibfnamefont {Y.}~\bibnamefont {Liu}}, \bibinfo {author}
  {\bibfnamefont {C.}~\bibnamefont {Stan}}, \bibinfo {author} {\bibfnamefont
  {W.}~\bibnamefont {Setiawan}}, \bibinfo {author} {\bibfnamefont
  {C.}~\bibnamefont {Sanner}}, \bibinfo {author} {\bibfnamefont
  {K.}~\bibnamefont {Xu}}, \ and\ \bibinfo {author} {\bibfnamefont
  {W.}~\bibnamefont {Ketterle}},\ }\href {\doibase 10.1038/nature05224}
  {\bibfield  {journal} {\bibinfo  {journal} {Nature}\ }\textbf {\bibinfo
  {volume} {443}},\ \bibinfo {pages} {961} (\bibinfo {year}
  {2006})}\BibitemShut {NoStop}%
\bibitem [{\citenamefont {Mitra}\ \emph {et~al.}(2017)\citenamefont {Mitra},
  \citenamefont {Brown}, \citenamefont {Guardado-Sanchez}, \citenamefont
  {Kondov}, \citenamefont {Devakul}, \citenamefont {Huse}, \citenamefont
  {Schaub},\ and\ \citenamefont {Bakr}}]{Mitra2017}%
  \BibitemOpen
  \bibfield  {author} {\bibinfo {author} {\bibfnamefont {D.}~\bibnamefont
  {Mitra}}, \bibinfo {author} {\bibfnamefont {P.~T.}\ \bibnamefont {Brown}},
  \bibinfo {author} {\bibfnamefont {E.}~\bibnamefont {Guardado-Sanchez}},
  \bibinfo {author} {\bibfnamefont {S.~S.}\ \bibnamefont {Kondov}}, \bibinfo
  {author} {\bibfnamefont {T.}~\bibnamefont {Devakul}}, \bibinfo {author}
  {\bibfnamefont {D.~A.}\ \bibnamefont {Huse}}, \bibinfo {author}
  {\bibfnamefont {P.}~\bibnamefont {Schaub}}, \ and\ \bibinfo {author}
  {\bibfnamefont {W.~S.}\ \bibnamefont {Bakr}},\ }\href {\doibase
  10.1038/nphys4297} {\bibfield  {journal} {\bibinfo  {journal} {Nature
  Physics}\ }\textbf {\bibinfo {volume} {14}},\ \bibinfo {pages} {173}
  (\bibinfo {year} {2017})}\BibitemShut {NoStop}%
\bibitem [{\citenamefont {Classen}\ \emph {et~al.}(2015)\citenamefont
  {Classen}, \citenamefont {Herbut}, \citenamefont {Janssen},\ and\
  \citenamefont {Scherer}}]{Classen2015}%
  \BibitemOpen
  \bibfield  {author} {\bibinfo {author} {\bibfnamefont {L.}~\bibnamefont
  {Classen}}, \bibinfo {author} {\bibfnamefont {I.~F.}\ \bibnamefont {Herbut}},
  \bibinfo {author} {\bibfnamefont {L.}~\bibnamefont {Janssen}}, \ and\
  \bibinfo {author} {\bibfnamefont {M.~M.}\ \bibnamefont {Scherer}},\ }\href
  {\doibase 10.1103/PhysRevB.92.035429} {\bibfield  {journal} {\bibinfo
  {journal} {Phys. Rev. B}\ }\textbf {\bibinfo {volume} {92}},\ \bibinfo
  {pages} {035429} (\bibinfo {year} {2015})}\BibitemShut {NoStop}%
\bibitem [{\citenamefont {Roy}\ \emph {et~al.}(2016)\citenamefont {Roy},
  \citenamefont {Juricic},\ and\ \citenamefont {Herbut}}]{Roy2016}%
  \BibitemOpen
  \bibfield  {author} {\bibinfo {author} {\bibfnamefont {B.}~\bibnamefont
  {Roy}}, \bibinfo {author} {\bibfnamefont {V.}~\bibnamefont {Juricic}}, \ and\
  \bibinfo {author} {\bibfnamefont {I.}~\bibnamefont {Herbut}},\ }\href
  {\doibase 10.1007/JHEP04(2016)018} {\bibfield  {journal} {\bibinfo  {journal}
  {J. High Energy Phys.}\ }\textbf {\bibinfo {volume} {04}},\ \bibinfo {pages}
  {018} (\bibinfo {year} {2016})}\BibitemShut {NoStop}%
\bibitem [{\citenamefont {Sato}\ \emph {et~al.}(2017)\citenamefont {Sato},
  \citenamefont {Hohenadler},\ and\ \citenamefont {Assaad}}]{Assaad2017}%
  \BibitemOpen
  \bibfield  {author} {\bibinfo {author} {\bibfnamefont {T.}~\bibnamefont
  {Sato}}, \bibinfo {author} {\bibfnamefont {M.}~\bibnamefont {Hohenadler}}, \
  and\ \bibinfo {author} {\bibfnamefont {F.~F.}\ \bibnamefont {Assaad}},\
  }\href {\doibase 10.1103/PhysRevLett.119.197203} {\bibfield  {journal}
  {\bibinfo  {journal} {Phys. Rev. Lett.}\ }\textbf {\bibinfo {volume} {119}},\
  \bibinfo {pages} {197203} (\bibinfo {year} {2017})}\BibitemShut {NoStop}%
\bibitem [{\citenamefont {Janssen}\ \emph {et~al.}(2018)\citenamefont
  {Janssen}, \citenamefont {Herbut},\ and\ \citenamefont
  {Scherer}}]{Janssen2018}%
  \BibitemOpen
  \bibfield  {author} {\bibinfo {author} {\bibfnamefont {L.}~\bibnamefont
  {Janssen}}, \bibinfo {author} {\bibfnamefont {I.~F.}\ \bibnamefont {Herbut}},
  \ and\ \bibinfo {author} {\bibfnamefont {M.~M.}\ \bibnamefont {Scherer}},\
  }\href {\doibase 10.1103/PhysRevB.97.041117} {\bibfield  {journal} {\bibinfo
  {journal} {Phys. Rev. B}\ }\textbf {\bibinfo {volume} {97}},\ \bibinfo
  {pages} {041117} (\bibinfo {year} {2018})}\BibitemShut {NoStop}%
\bibitem [{\citenamefont {Torres}\ \emph {et~al.}(2020)\citenamefont {Torres},
  \citenamefont {Weber}, \citenamefont {Janssen}, \citenamefont {Wessel},\ and\
  \citenamefont {Scherer}}]{Torres2020}%
  \BibitemOpen
  \bibfield  {author} {\bibinfo {author} {\bibfnamefont {E.}~\bibnamefont
  {Torres}}, \bibinfo {author} {\bibfnamefont {L.}~\bibnamefont {Weber}},
  \bibinfo {author} {\bibfnamefont {L.}~\bibnamefont {Janssen}}, \bibinfo
  {author} {\bibfnamefont {S.}~\bibnamefont {Wessel}}, \ and\ \bibinfo {author}
  {\bibfnamefont {M.~M.}\ \bibnamefont {Scherer}},\ }\href {\doibase
  10.1103/PhysRevResearch.2.022005} {\bibfield  {journal} {\bibinfo  {journal}
  {Phys. Rev. Research}\ }\textbf {\bibinfo {volume} {2}},\ \bibinfo {pages}
  {022005} (\bibinfo {year} {2020})}\BibitemShut {NoStop}%
\bibitem [{\citenamefont {Shirakawa}\ \emph {et~al.}(2020)\citenamefont
  {Shirakawa}, \citenamefont {Miyakoshi},\ and\ \citenamefont
  {Yunoki}}]{Shirakawa2020}%
  \BibitemOpen
  \bibfield  {author} {\bibinfo {author} {\bibfnamefont {T.}~\bibnamefont
  {Shirakawa}}, \bibinfo {author} {\bibfnamefont {S.}~\bibnamefont
  {Miyakoshi}}, \ and\ \bibinfo {author} {\bibfnamefont {S.}~\bibnamefont
  {Yunoki}},\ }\href {\doibase 10.1103/PhysRevB.101.174307} {\bibfield
  {journal} {\bibinfo  {journal} {Phys. Rev. B}\ }\textbf {\bibinfo {volume}
  {101}},\ \bibinfo {pages} {174307} (\bibinfo {year} {2020})}\BibitemShut
  {NoStop}%
\bibitem [{\citenamefont {You}\ \emph {et~al.}(2017)\citenamefont {You},
  \citenamefont {Ludwig},\ and\ \citenamefont {Xu}}]{You2017}%
  \BibitemOpen
  \bibfield  {author} {\bibinfo {author} {\bibfnamefont {Y.-Z.}\ \bibnamefont
  {You}}, \bibinfo {author} {\bibfnamefont {A.~W.~W.}\ \bibnamefont {Ludwig}},
  \ and\ \bibinfo {author} {\bibfnamefont {C.}~\bibnamefont {Xu}},\ }\href
  {\doibase 10.1103/PhysRevB.95.115150} {\bibfield  {journal} {\bibinfo
  {journal} {Phys. Rev. B}\ }\textbf {\bibinfo {volume} {95}},\ \bibinfo
  {pages} {115150} (\bibinfo {year} {2017})}\BibitemShut {NoStop}%
\end{thebibliography}%

\end{document}